\documentclass[a4paper,fleqn,usenatbib]{mnras}

\usepackage{newtxtext,newtxmath}

\usepackage[T1]{fontenc}
\usepackage{ae,aecompl}
\usepackage{url}
\usepackage{bm}
\usepackage{soul}

\usepackage{etoolbox}
\makeatletter
\patchcmd\@combinedblfloats{\box\@outputbox}{\unvbox\@outputbox}{}{%
  \errmessage{\noexpand\@combinedblfloats could not be patched}%
}%
\makeatother

\usepackage{graphicx}	
\usepackage{amsmath}	
\usepackage{amssymb}	

\newcommand{\SFRu} {$ \,{\rm M}_{\odot}/{\rm yr}$~}
\newcommand{\chone}{$[3.6]$~}
\newcommand{\chtwo}{$[4.5]$~}
\newcommand{\oiii}{{\sc [OIII]}~}

\title[Dust in bright $z \sim 7$ LBGs]{Obscured star-formation in bright \boldmath$z \simeq 7$ Lyman-break galaxies} 

\author[R.A.A.Bowler et al.]{
R. A. A. Bowler,$^{1}$\thanks{E-mail: rebecca.bowler@physics.ox.ac.uk}
N. Bourne,$^{2}$
J. S. Dunlop,$^{2}$
R. J. McLure$^{2}$ and
D. J. McLeod$^{2}$
\\
$^{1}$Astrophysics, The Denys Wilkinson Building, University of Oxford, Keble Road, Oxford, OX1 3RH \\
$^{2}$Institute for Astronomy, University of Edinburgh, Royal Observatory, Edinburgh, EH9 3HJ\\
}

\date{Accepted XXX. Received YYY; in original form ZZZ}

\pubyear{2018}

\begin{document}
\label{firstpage}
\pagerange{\pageref{firstpage}--\pageref{lastpage}}
\maketitle

\begin{abstract}
We present Atacama Large Millimeter/Submillimeter Array observations of the rest-frame far-infrared (FIR) dust continuum emission of six bright Lyman-break galaxies (LBGs) at $z \simeq 7$.
One LBG is detected ($5.2\sigma$ at peak emission), while the others remain individually undetected at the $3\sigma$ level.
The average FIR luminosity of the sample is found to be $L_{\rm FIR} \simeq 2 \times 10^{11}\,{\rm L}_{\odot}$, corresponding to an obscured star-formation rate (SFR) that is comparable to that inferred from the unobscured UV emission.
In comparison to the infrared excess (IRX$\,=L_{\rm FIR}/L_{\rm UV}$)-$\beta$ relation, our results are consistent with a Calzetti-like attenuation law (assuming a dust temperature of T$ = 40$--$50$K).
We find a physical offset of $3\,{\rm kpc}$ between the dust continuum emission and the rest-frame UV light probed by~\emph{Hubble Space Telescope} imaging for galaxy ID65666 at $z = 7.17^{+0.09}_{-0.06}$.
The offset is suggestive of an inhomogeneous dust distribution, where  $75$\% of the total star formation activity (SFR$ \,\simeq 70\,{\rm M}_{\odot}/{\rm yr}$) of the galaxy is completely obscured. 
Our results provide direct evidence that dust obscuration plays a key role in shaping the bright-end of the observed rest-frame UV luminosity function at $z \simeq 7$, in agreement with cosmological galaxy formation simulations.
The existence of a heavily-obscured component of galaxy ID65666 indicates that dusty star-forming regions, or even entire galaxies, that are ``UV-dark'' are significant even in the $z \simeq 7$ galaxy population.

\end{abstract}




\section{Introduction}
The onset of significant dust creation and obscuration in the Universe remains a key question in the study of early galaxy formation.
While observations and theoretical predictions indicate that dust production can be rapid (e.g.~\citealp{Indebetouw2014, Todini2001}), the relative importance of dust creation in supernovae and Asymptotic Giant-branch stars, grain growth, and dust destruction at high redshifts remains uncertain (e.g.~\citealp{Valiante2009, Popping2017, Michalowski2015}).
The first direct observations of the far-infrared (FIR) dust continuum at the highest redshifts were in extremely dusty star-forming systems (e.g.~\citealp{Casey2014, Riechers2013}) or the host-galaxies of luminous quasars (e.g.~\citealp{Bertoldi2003, Wang2007}).
Now, with the sensitivity provided by the Atacama Large Millimeter/Submillimeter Array (ALMA), it is possible to measure the FIR continuum in the more `typical' star-forming galaxy population at $z \gtrsim 6$ for the first time~\citep{Capak2015, Watson2015, Riechers2014}.

Locally, the FIR emission of starburst galaxies has been found to correlate with the rest-frame UV slope ($\beta$, $F_{\lambda} \propto \lambda^\beta$) through the `infrared-excess-$\beta$' relation (IRX = $L_{\rm FIR}/L_{\rm UV}$:~\citealp{Meurer1999}).
Such a relationship has been found to hold up to $z \simeq 3$ (e.g.~\citealp{McLure2017, Fudamoto2017, Reddy2018, Bourne2017, AlvarezMarquez2016}), with several studies finding a lower normalisation that is suggestive of a steeper Small Magellanic Cloud (SMC)-like dust law (\citealp{Reddy2018, Bouwens2016b}; although see~\citealp{McLure2017, Koprowski2018, Bourne2017} who found consistency with the local relation).
At $z > 5$ however, the form or even existence of an IRX-$\beta$ relation is uncertain.
The first study to look at a sample of Lyman-break galaxies at $z = 5$--$6$ with ALMA found a deficit of FIR emission than expected from the local IRX-$\beta$ relations~\citep{Capak2015, Willott2015}, with several galaxies in the~\citet{Capak2015} sample being undetected in the FIR continuum with ALMA despite showing red rest-frame UV colours.
A stacking analysis of the ALMA Spectroscopic Survey data in the Hubble Ultra Deep Field by~\citet{Bouwens2016b} also found fewer detections at $z > 4$ than expected assuming the local IRX-$\beta$ relation.
These initial studies were interpreted as evidence for different interstellar medium properties in LBGs at high redshift.
More recently however,~\citet{Koprowski2018} and~\citet{McLure2017} have identified that the derived IRX-$\beta$ obtained through stacking in bins of $\beta$ (such as that presented in~\citealp{Bouwens2016b}) are biased to low IRX values.
Instead,~\citet{Koprowski2018} find evidence that a Calzetti-like attenuation law well describes the observations of LBGs up to $z = 4.8$.

In addition to theses results,  there have several direct detections of FIR continuum emission from individual LBGs at very high redshifts.
The gravitationally lensed galaxy A1689-zD1 at $z = 7.5$~\citep{Watson2015}, which has a blue rest-frame UV colour ($\beta \simeq -2$) suggestive of little dust attenuation, lies significantly in excess of the local IRX-$\beta$ relationship and the results of~\citet{Capak2015} and~\citet{Bouwens2016b}.
\citet{Laporte2017} also claim a detection of the dust continuum in a $z = 8.38$ LBG, which has comparable FIR luminosity to A1689-zD1.
The resulting picture is one of significant scatter in the IRX at a given $\beta$ slope, with no compelling evidence for a correlation in the IRX-$\beta$ plane at $z > 5$.
Further uncertainties are introduced by the unknown dust temperature and FIR spectral-energy distribution (SED) of high-redshift LBGs~\citep{Faisst2017} and the difficulty in measuring reliable $\beta$-slopes with typically faint detections in the observed near-infrared~\citep{Barisic2017}.

Amongst UV-selected LBGs there exists a colour-magnitude relationship until at least $z = 5$~\citep{Rogers2014, Bouwens2014beta}, with the most UV luminous galaxies tending to appear redder on average, with the implication that the most luminous galaxies suffer increased dust obscuration.
It is therefore expected that the brightest galaxies should be dustier than the fainter galaxy population at this epoch.
Indeed, at $z \simeq 5$,~\citet{Cullen2017} have shown that a luminosity dependent dust correction motivated by the colour-magnitude relation can reproduce the observed rest-frame UV luminosity function (LF) in the First Billion Years (FiBY) simulation.
In practice, many of the leading cosmological galaxy formation simulations and models currently include significant dust attenuation in order to match the observed rest-frame UV LFs at high redshifts.
For example, Illustris~\citep{Genel2014}, the Munich models~\citep{Henriques2014} and the analytical model of~\citet{Cai2014} all apply dust obscuration of the order of $A_{\rm UV} \simeq 1$--$2\,$mag to match the observed number densities of the brightest galaxies at $z \simeq 7$ (see fig. 13 in~\citealp{Bowler2015}).
What is required to test such models are observations that directly probe the emission from dust and hence place independent constraints on the attenuation in the rest-frame UV.
In particular, observations of the brightest LBGs are a promising avenue of study as they are 1) predicted to be the dustiest galaxies from the colour-magnitude relation and 2) can be studied in detail with modest telescope time.
Such observations are essential to determine the importance of dust obscuration (as opposed to other effects such as mass quenching;~\citealp{Peng2010}) in shaping the bright-end of the UV LF at the highest redshifts.

In this study we present the results of a targeted ALMA follow-up programme to image six bright $z \simeq 7$ LBGs discovered within $1\,{\rm deg}^2$ of overlapping deep optical/near-infrared data in the Cosmological Evolution Survey (COSMOS) field~\citep{Bowler2012, Bowler2014}. 
We describe the new ALMA dataset and auxiliary imaging in Section~\ref{sect:data}.
The results are presented in Section~\ref{sect:results}, which is followed by a comparison to the IRX-$\beta$ relation in Section~\ref{sect:irx}.
A discussion of the physical offset found between the observed dust and UV emission for ID65666, SED fitting results and implications for galaxy evolution simulations is presented is Section~\ref{sect:diss}.
We end with our conclusions in Section~\ref{sect:conclusions}.
We assume a cosmology with $H_{0} = 70\,{\rm km}{\rm s}^{-1}\,{\rm Mpc}^{-1}$, $\Omega_{\rm m} = 0.3$ and $\Omega_\Lambda = 0.7$.
Magnitudes throughout are presented in the AB system~\citep{Oke1974, Oke1983}.

\section{Data \& Sample}\label{sect:data}

\begin{table}
\caption{The coordinates of the six galaxies observed with ALMA.  The coordinates correspond to the ground-based UltraVISTA DR3 centroid.  The photometric redshift is shown in the final column.  Furthermore, UVISTA-65666 has been spectroscopically confirmed at $z = 7.152$~\citep{Hashimoto2018}.}
\label{table:flux}
\begin{tabular}{l l c c}

\hline
ID & R.A.(J2000) & Dec.(J2000) & $z_{\rm phot}$   \\
\hline

UVISTA-65666 & 10:01:40.69 & +01:54:52.42 & $7.17_{-0.06}^{+0.09}$\\[1ex]
UVISTA-304416 & 10:00:43.38 & +02:37:51.73 & $6.97_{-0.04}^{+0.08}$\\[1ex]
UVISTA-238225 & 10:01:52.31 & +02:25:42.24 & $6.88_{-0.08}^{+0.08}$\\[1ex]
UVISTA-169850 & 10:02:06.47 & +02:13:24.19 & $6.68_{-0.04}^{+0.05}$\\[1ex]
UVISTA-304384 & 10:01:36.86 & +02:37:49.14 & $6.66_{-0.05}^{+0.06}$\\[1ex]
UVISTA-279127 & 10:01:58.51 & +02:33:08.58 & $6.58_{-0.05}^{+0.04}$\\

\hline

\end{tabular}
\end{table}

The ALMA data was obtained in Cycle 3 through proposal 2015.1.00540.S (PI Bowler).
We observed the six brightest Lyman-break galaxies at $6.5 < z < 7.5$ presented in~\citet{Bowler2012, Bowler2014}.
The galaxies are all within the COSMOS field~\citep{Scoville2007}\footnote{\url{http://cosmos.astro.caltech.edu}}, and were initially selected in the near-infrared $YJHK_{s}$ data based on the UltraVISTA\footnote{\url{http://ultravista.org/}} survey data release 2~\citep{McCracken2012}.
The coordinates of the six galaxies, which defined each ALMA pointing, and their photometric redshifts are shown in Table~\ref{table:flux}.
The galaxies were selected via SED fitting and are secure $z > 6.5$ objects as demonstrated by the sharp spectral break observed in the photometric data at $\simeq 1\,\mu{\rm m}$ (Fig~\ref{fig:sed}).
The strength of the break ($z - Y > 2\,{\rm mag}$) observed in the sample cannot be reproduced by the common low-redshift or brown-dwarf contaminants (e.g.~\citealp{Ouchi2010, Bouwens2011,Bowler2012}).
Each target was observed for $10$ minutes in Band 6 (centred at $233\,{\rm GHz}$ or $1.29\,{\rm mm}$; corresponding to $\sim 170\mu {\rm m}$ in the rest-frame).
The integration time was chosen to provide limits on the ratio of FIR to UV luminosity sufficient to distinguish between the previous results at $z > 5$ from~\citet{Watson2015} and~\citet{Capak2015}.
The data was taken in a compact configuration C36-2, with a maximum baseline of 330m, on 25 January 2016.
In order to maximise continuum sensitivity, the correlator was set up in time division mode with 2~GHz bandwidth across each of four spectral windows (centred at 224, 226, 240, 242 GHz) in two polarizations.
The data was reduced with the ALMA pipeline of \textsc{casa} version 4.5.1.
Calibration was performed by ALMA personnel using J1058+0133, J0948+0022 and Ganymede for bandpass, phase and flux calibration respectively. We then processed the data through the imaging pipeline, collapsing the full 8 GHz bandwidth and both polarizations into a single continuum image of each field, initially using natural weighting of baselines to obtain the optimal root-mean-square (RMS) sensitivity for point sources. 
Imaging was achieved using the \textsc{tclean} task, and the data were cleaned by defining a mask for any source in the field with a peak signal-to-noise ratio SNR~$>3$ that also corresponds to a~\emph{HST}-detected galaxy (whether the target or serendipitous), with a clean threshold of 60 $\mu$Jy ($\simeq 2\times$~RMS).
The natural weighted images have a beam size of $1.1 \times 1.4$ arcsec full-width at half-maximum (FWHM) and reach an average RMS sensitivity of $27\, \mu {\rm Jy}/{\rm beam}$, with small deviations of $2 \, \mu {\rm Jy}/{\rm beam}$ between the different pointings.
To increase the sensitivity to extended flux, we also tried imaging the targets with a 1 arcsec taper, which results in a beam size of $1.4 \times 1.7$ arcsec and RMS sensitivity of $31\, \mu {\rm Jy}/{\rm beam}$.
As the natural weighted images have a smaller beam and better RMS sensitivity, we chose to present this weighting throughout the paper, however comment on the tapered results where relevant.
We measured fluxes using two complementary methods.
First, fluxes were measured from the peak flux within an aperture of radius $1\,$arcsec from the~\emph{HST}/WFC3 centroid.
The peak flux assumes that the source is unresolved in the ALMA data.
For the stacks and the detected individual source, we also measured the total flux using a Gaussian fit.
In the case of a non-detection at the $3\sigma$ level for the individual LBGs, we conservatively provide $2\sigma$ upper limits calculated as $f_{\rm peak} + 2\,\times\,{\rm RMS}$.

\begin{figure}

\includegraphics[width = 0.23\textwidth]{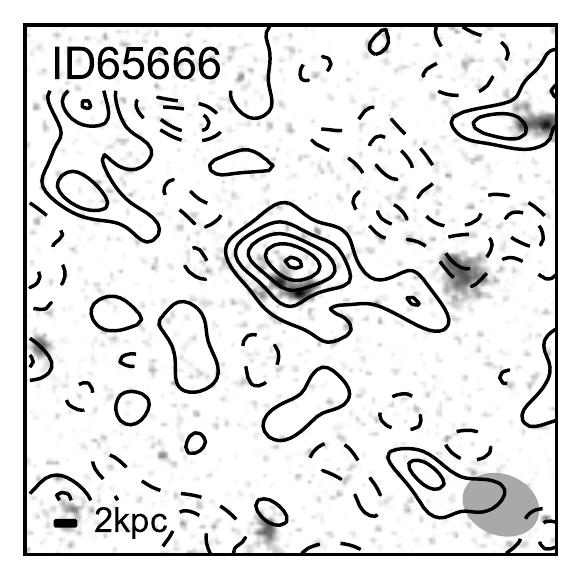} 
\includegraphics[width = 0.23\textwidth]{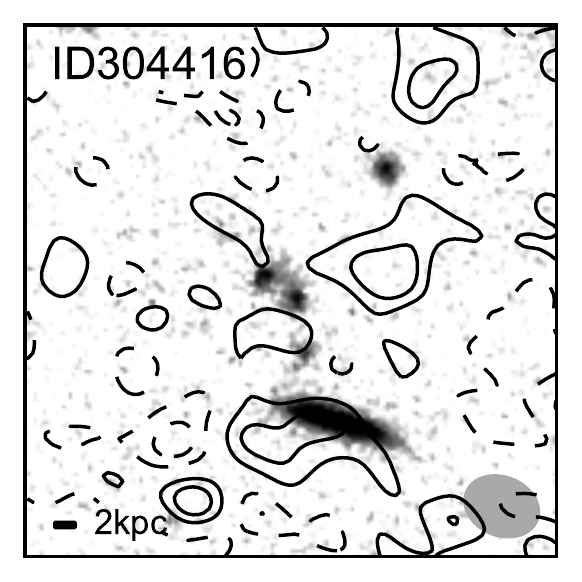}\\
\includegraphics[width = 0.23\textwidth]{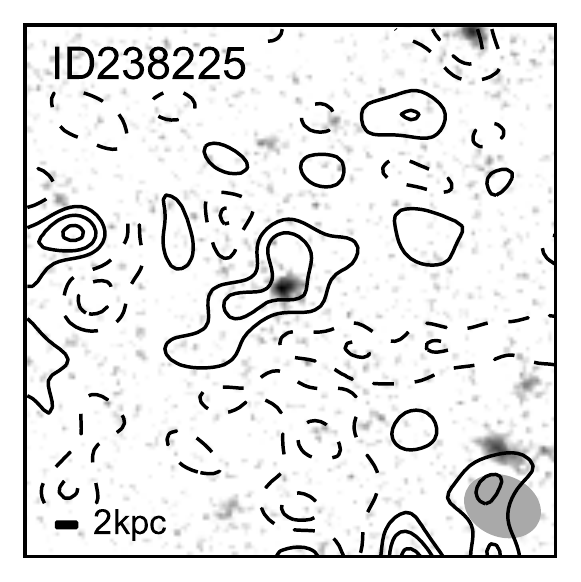}
\includegraphics[width = 0.23\textwidth]{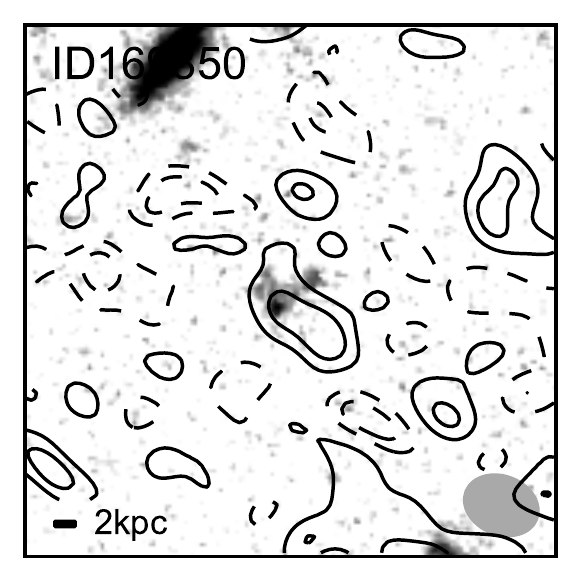} \\
\includegraphics[width = 0.23\textwidth]{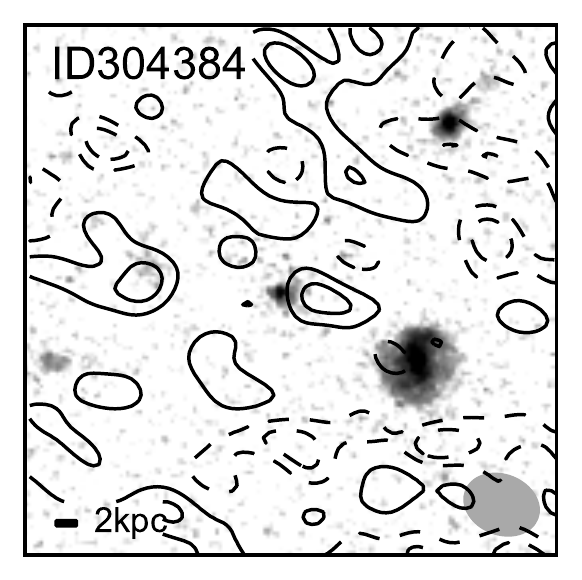} 
\includegraphics[width = 0.23\textwidth]{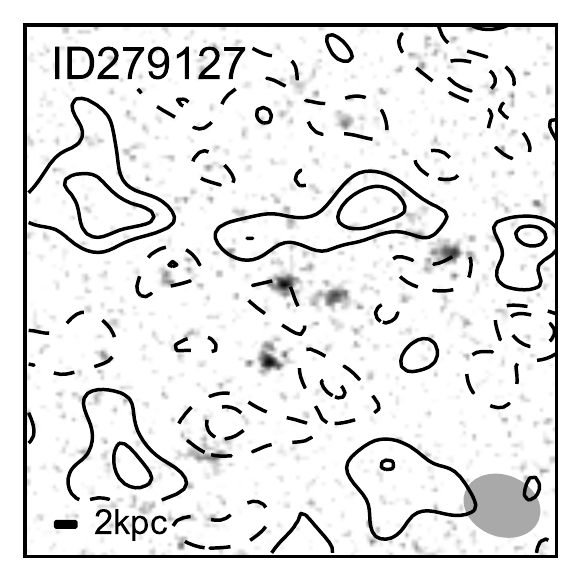} \\

\caption{The six $z \simeq 7$ Lyman-break galaxies targeted in this study, shown in order of descending redshift from top left to lower right.
The background image is the~\emph{HST}/WFC3 $JH_{140}$ data, and the contours show the results of our ALMA $1.3\,{\rm mm}$ observations with natural weighting.
The images are $10$ arcsec on the side, with North to the top and East to the left.
The ALMA contours are drawn at $1 \, \sigma$ intervals (negative contours are shown as dotted lines).
The beam is shown as the ellipse in the lower right-hand corner of each stamp.
The greyscale for the background~\emph{HST} image is linearly scaled from a minimum surface brightness of $26\,{\rm mag}/{\rm arcsec}^2$ to the peak brightness of the primary target.}\label{fig:stamps}

\end{figure}

High-resolution rest-frame UV imaging of the sample is available from~\emph{HST}/Wide Field Camera 3 (WFC3) as described in~\citet{Bowler2016}.
The imaging consists of a single orbit in the $JH_{140}$ filter, which reaches a $5\sigma$ depth of $m_{\rm AB} = 27.0$ (0.6 arcsec diameter aperture).
For this sample the $JH_{140}$ filter is uncontaminated by any Lyman-$\alpha$ emission, and hence provides a direct measurement of the rest-frame UV continuum at $\sim 1800\,$\AA.
The galaxy photometric redshifts and stellar masses were determined from SED fitting to the deep multi-wavelength photometric data available over the COSMOS field.
This includes~\emph{Spitzer}/IRAC data from the SPLASH~\citep{Steinhardt2014}\footnote{\url{http://splash.caltech.edu}} and SMUVS~\citep{Ashby2018} surveys, near-infrared data from the third data release of UltraVISTA~\citep{McCracken2012} and finally optical data from Subaru/SuprimeCam~\citep{Furusawa2016} and the Hyper-SurprimeCam (HSC) Subaru Strategic Program~\citep{Aihara2017}.
The photometric redshift code {\sc LePhare} was used~\citep{Arnouts1999, Ilbert2006} to perform the SED fitting analysis.
We fitted~\citet{Bruzual2003} models assuming an exponentially declining star-formation history (SFH), a~\citet{Chabrier2003} initial-mass function, and a metallicity of $1/5\,{\rm Z}_{\odot}$.
The $\tau$ parameter defining the characteristic timescale of the SFH ranged between $50\,{\rm Myr}$ and $10\,{\rm Gyr}$, approximating both a burst and constant SFH in the extreme cases.
The minimum galaxy age was set to $10\,{\rm Myr}$, with the maximum age set by the age of the Universe at each redshift.
Dust attenuation was applied assuming a~\citet{Calzetti2000} dust law.
The best-fitting SED models, along with the photometry for our sample is shown in Fig.~\ref{fig:sed}.
We measured photometry in $1.8''$ diameter apertures, correcting to total flux using a point source correction determined using {\sc PSFEx}~\citep{Bertin2011}.
Errors were obtained using local depths determined from the closest 200 empty apertures to each galaxy.
\emph{Spitzer}/IRAC data was deconfused using {\sc tphot} following the procedure outlined in~\citet{McLeod2016}.
Further details of the ground-based images and data reduction can be found in~\citet{Bowler2014}.

\section{Results} \label{sect:results}

\begin{table*}
\caption{The observed ALMA flux and derived properties of the sample of six bright LBGs.
The galaxies are ordered in descending redshift, as in Table~\ref{table:flux}.
The upper part of the table shows the individual results, and the lower part shows the stacked results.
The rest-frame UV slope, ($\beta$; see Section~\ref{sect:irx}) is presented in column 2, followed by the absolute UV magnitude and the corresponding monochromatic luminosity in columns 3 and 4.
The observed 1.3mm flux, or upper limit on the flux, is shown in column 5.
For the detected object ID65666, the total flux is determined from a Gaussian fit.
This is followed by the FIR luminosity derived using a modified blackbody fit assuming $T = 40\,$K is shown in column 6.
The ratio of the FIR to UV luminosity, IRX, is presented in column 7.
Finally, in columns 8 and 9 we show the SFR in the FIR and the stellar mass respectively.
}\label{table:properties}
\begin{tabular}{l c c c c c c c c c c}
\hline
ID & $\beta$ & $M_{\rm UV}$ & $L_{\rm UV}$ & $f_{\rm1.3mm}$ & $L_{\rm FIR}$ & ${\rm log}({\rm IRX})$ & SFR$_{\rm FIR}$ & ${\rm log}(M_*/{\rm M}_{\odot})$  \\
& & (mag) & $ (10^{11} {\rm L}_{\odot}$) & ($\mu{\rm Jy}$) & ($ 10^{11} {\rm L}_{\odot}$) &  & (\SFRu) &  & \\

\hline

UVISTA-65666 & $-1.85_{-0.53}^{+0.54}$ & $ -22.43_{-0.10}^{+0.11} $ & $ 1.98 \pm 0.19 $ & $168 \pm 56$ & $ 3.63 \pm 1.21 $ & $ \phantom{-}0.26_{-0.22}^{+0.17} $ & $ 50\pm17$ & $ 9.0_{-0.2}^{+0.8
}$\\[1ex]
UVISTA-304416 & $-2.26_{-0.53}^{+0.53}$ & $ -23.16_{-0.09}^{+0.10} $ & $ 3.89 \pm 0.34 $ & $< 106$ & $ <2.36 $ & $ <-0.22 $ & $ < 33$ & $ 9.6_{-0.2}^{+0.4}$\\[1ex]
UVISTA-238225 & $-2.00_{-0.63}^{+0.59}$ & $ -22.41_{-0.12}^{+0.12} $ & $ 1.95 \pm 0.22 $ & $< 139$ & $ <3.02 $ & $ <0.19 $ & $ < 42$ & $ 9.6_{-0.2}^{+0.4}$\\[1ex]
UVISTA-169850 & $-1.80_{-0.29}^{+0.29}$ & $ -22.92_{-0.10}^{+0.12} $ & $ 3.12 \pm 0.31 $ & $< 129$ & $ <2.90 $ & $ <-0.03 $ & $ < 40$ & $ 9.4_{-0.2}^{+0.6}$\\[1ex]
UVISTA-304384 & $-1.86_{-0.36}^{+0.35}$ & $ -22.40_{-0.10}^{+0.11} $ & $ 1.93 \pm 0.19 $ & $< 127$ & $ <2.84 $ & $ <0.17 $ & $ < 39$ & $ 9.5_{-0.2}^{+0.6}$\\[1ex]
UVISTA-279127 & $-2.25_{-0.39}^{+0.37}$ & $ -22.62_{-0.15}^{+0.18} $ & $ 2.36 \pm 0.36 $ & $< 92$ & $ <2.00 $ & $ <-0.07 $ & $ < 28$ & $ 9.5_{-0.6}^{+0.6}$\\
\hline
Full stack & $-2.00_{-0.19}^{+0.19}$ & $ -22.76_{-0.06}^{+0.05} $ & $ 2.69 \pm 0.14 $ & $101 \pm 34$ & $ 2.26 \pm 0.76 $ & $ -0.08_{-0.20}^{+0.15} $ & $ 31\pm11$ & --\\[1ex]
Non detections & $-2.03_{-0.20}^{+0.20}$ & $ -22.81_{-0.06}^{+0.05} $ & $ 2.82 \pm 0.14 $ & $100 \pm 50$ & $ 2.23 \pm 1.12 $ & $ -0.10_{-0.32}^{+0.20} $ & $ 31\pm16$ & --\\

\hline
\end{tabular}
\end{table*}

The new ALMA data for our sample is shown in comparison to the~\emph{HST}/WFC3 $JH_{140}$ imaging in Fig.~\ref{fig:stamps}.
We find that one galaxy in our sample (ID65666) is clearly detected in the ALMA images, with a measured peak flux of $124 \pm 24\,\mu{\rm Jy}$ or $5.2\sigma$ significance.
ID65666 is the highest-redshift galaxy we targeted, with a best-fitting photometric redshift of $z = 7.17_{-0.06}^{+0.09}$.
The inclusion of new HSC data as shown in Fig.~\ref{fig:sed}, results in a slightly higher (but consistent) photometric redshift than previously obtained ($z = 7.04_{-0.15}^{+0.16}$;~\citealp{Bowler2016}).
The photometric redshift we find is consistent with the tentative detection of the Lyman-$\alpha$ emission line at $z = 7.168$ for this object by~\citet{Furusawa2016}, and the recently  the spectroscopic redshift of $z = 7.152$ derived from the FIR lines {\sc [OIII]} $88\mu{\rm m}$ and {\sc [CII]} $158\mu{\rm m}$ by~\citet{Hashimoto2018}.
We found a slight ($10$\%) increase in signal-to-noise for ID65666 when using the $1$ arcsec tapered image, indicating that the dust emission is extended on this scale.
Such an extension is comparable to the size of the rest-frame UV emission seen in Fig.~\ref{fig:stamps}.
Using a Gaussian fit to account for this extended emission results in a total flux of $168 \pm 56\,\mu\,{\rm Jy}$.
The ALMA detection and the $JH_{140}$ image show an offset of $\simeq 0.6\,$ arcsec, which we discuss further in Section~\ref{sect:offset}. 
The remaining galaxies are undetected at the $3\sigma$ level.
The measured fluxes and $2\sigma$ upper limits are presented in Table~\ref{table:properties}.
Four of the six LBGs presented here show multiple components in the high-resolution~\emph{HST}/WFC3 imaging.
In~\citet{Bowler2016} we checked~whether each clump was consistent with being at $z > 6.5$ using the deep~\emph{HST}/Advanced Camera for Surveys $I_{814}$ data available in the COSMOS field~\citep{Koekemoer2007,Scoville2007a, Massey2010}.
This analysis confirmed that the components found within 0.5 arcsec\footnote{At separations of greater than $\sim 0.5$ arcsec from the high-redshift target, we find lower redshift galaxies along the line of sight.  In particular, the small clump between the edge-on spiral and the dual cored LBG in object ID304416, is detected in the~\emph{HST}/ACS data and is therefore not associated with the $z = 6.85$ galaxy.
Furthermore, the object to the lower left of the two more widely separated components of ID279127 is also detected in the $I_{814}$ data and is therefore also at lower redshift.} were genuinely associated with the high-redshift LBG, with each clump showing a strong optical-to-near-infrared colour exceeding $I_{814} - JH_{140} \gtrsim 1.5$.
The probability of chance alignment of a very red, low-redshift interloper, with our high-redshift sample, is expected to be very small ($< 1$ percent;~\citealp{Bowler2016}).

The calculation of the total rest-frame FIR luminosity ($L_{\rm FIR}$) from the observed single data point at $1.3\,{\rm mm}$ requires the assumption of a spectral-energy distribution in the FIR.
To provide a straightforward comparison to previous studies we calculated the $L_{\rm FIR}$ first using a simple modified blackbody (greybody) curve, of the form $f_\nu \propto \nu^\beta B_\nu(T)$, assuming an optically thin medium.
We fix the exponent to $\beta = 1.6$, which is the best-fitting value from the sample of local infrared luminous galaxies presented in~\citep{Casey2012}.
A dust temperature of $T = 40\,$K was taken as our fiducial value.
This is motived by the results of~\citet{Coppin2015}, who found a best-fitting dust temperature of $T = 37$--$38\,$K through a stacking analysis of LBGs at $z = 3$--$5$ (see also~\citealt{Koprowski2018, Fudamoto2017, Knudsen2016}).
Assuming this dust temperature allows a more direct comparison to previous works from~\citet{Watson2015}, who assumed $40\,$K and to~\citet{Capak2015} who used dust temperatures in the range of $T = 25$--$45\,$K.
It has been suggested that at high-redshift the dust temperature could be higher than is typically found at lower redshifts (e.g.~\citealp{Schreiber2017, Faisst2017}).
Hence we also fit the observations assuming a higher dust temperature of $50\,$K.
We investigated the effect of allowing $\beta$ to vary according to the $1\sigma$ range found at lower redshift~\citep{Casey2012}.
The resulting variation in the derived FIR luminosity was found to be $<10$ percent.
As this additional error is small in comparison to the error on the flux, we neglect it in the following analysis.
In~\citet{Capak2015} an alternative parameterisation of the FIR SED was used that includes an additional power-law component in the mid-IR.
Using such a parameterisation results in brighter FIR luminosities by a factor of $1.63 \pm 0.15$ (using the form and best-fit parameters, and errors, presented by~\citealp{Casey2012}).
Finally, we calculate $L_{\rm FIR}$ using the empirical FIR SEDs from several lower-redshift starburst galaxies: Arp220, M82, and with the average SED of sub-mm galaxies derived by~\citet{Michalowski2010} and~\citet{Pope2008}.
We account for the reduced contrast due to the higher temperature of the Cosmic Microwave Background (CMB; $T\simeq 22\,$K at $z = 7$) using the approach presented in~\citet{DaCunha2013}.
The dust temperatures quoted above correspond to the temperature after any CMB heating has taken place.
The total luminosity was calculated by integrating under the FIR SED from $8\,\mu$m to $1000\,\mu$m.
The resulting $L_{\rm FIR}$ assuming a dust temperature of $T = 40\,$K are presented in Table~\ref{table:properties}.
The assumption of a higher dust temperature results in an increase in the FIR luminosity of a factor of $2$.
Using the range of empirical galaxy templates results in an increase of $\simeq 1$ for Apr220, $\simeq 2$ for M82 and $\simeq 0.9$ and $\simeq 1.5$ for the sub-mm templates of~\citet{Michalowski2010} and~\citet{Pope2008} respectively.

\begin{figure}

\includegraphics[width = 0.48\textwidth]{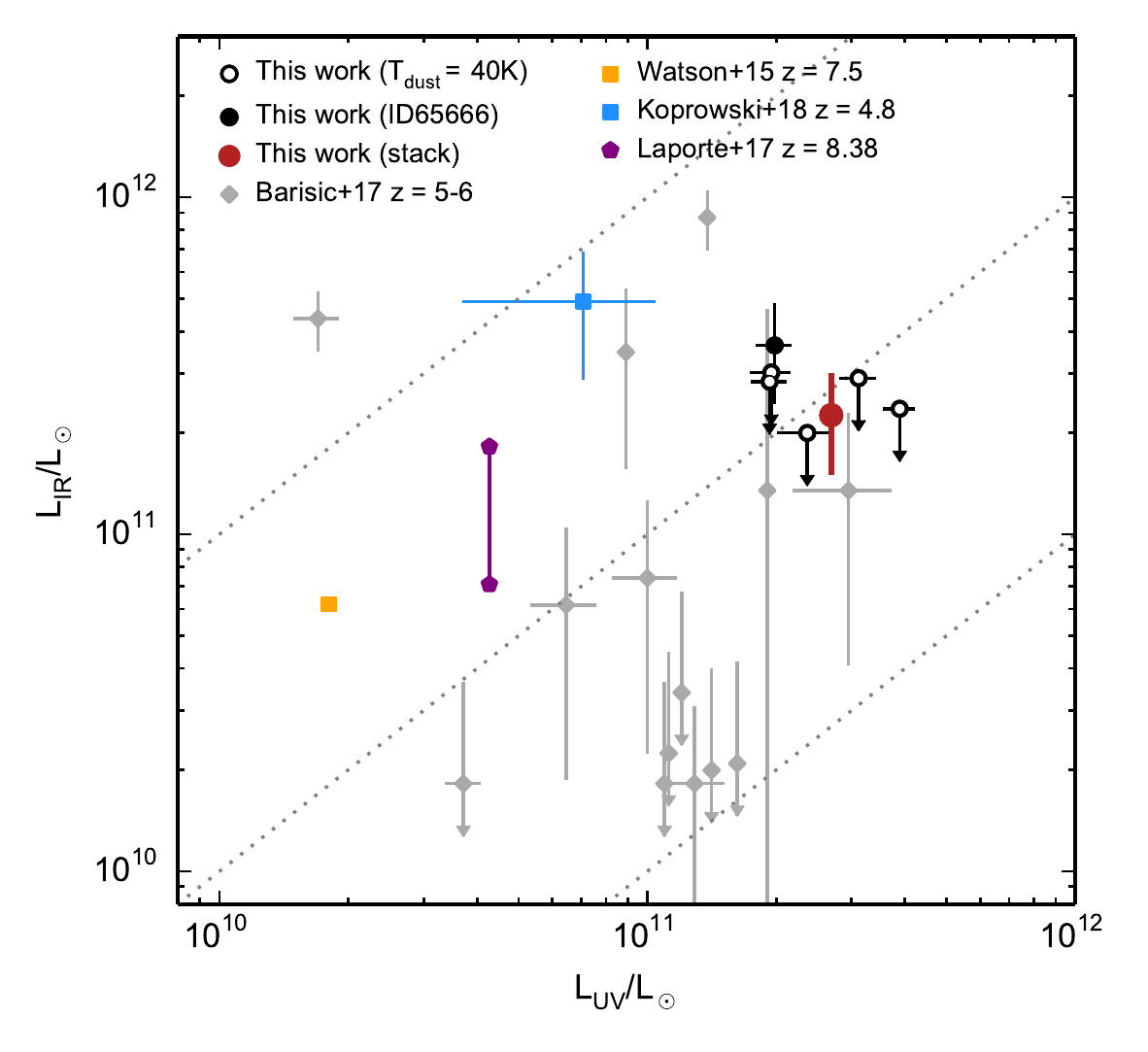}

\caption{The rest-frame UV luminosity compared to the derived rest-frame FIR luminosity for a compilation of studies at $z \gtrsim 5$, where ALMA detections have been found.
For~\citet{Laporte2017} we display the range of acceptable values derived in their work.
The dotted lines show the position of $IRX = [1.0, 0.0, -1.0]$ from top to bottom to guide the eye.  The LBGs that we targeted with ALMA represent the most UV luminous galaxies known at $z \simeq 7$ to-date.
}\label{fig:luvlir}

\end{figure}

The detection and upper limits on the FIR luminosity that we derive exceed that found in the majority of previous ALMA detections at $z > 5$.
We compare our derived UV and FIR luminosities for the sample to those from previous studies at these redshifts in Fig.~\ref{fig:luvlir}.
The detected galaxy, ID65666, shows $L_{\rm FIR} = 3.63 \pm 1.21 \times 10^{11}\,L_{\odot}$ ($T = 40\,$K) which is substantially brighter than 80\% of the sample presented by~\citet{Capak2015} at $z = 5$--$6$ LBGs.
ID65666 is over twice as luminous in the FIR than both of the $z > 7$ detections presented by~\citet{Watson2015} and~\citet{Laporte2017} ($L_{\rm FIR} = 6.2 \times 10^{10}\,L_{\odot}$ and $L_{\rm FIR} = 7$--$18 \times 10^{10}\,L_{\odot}$ respectively).
For the other five galaxies with non-detections in the ALMA data, we derive conservative upper limits of $L_{\rm FIR} \lesssim 2$--$3 \times 10^{11}\,L_{\odot}\,$($2\sigma$).

\subsection{Stacking results}

\begin{figure}

\includegraphics[width = 0.23\textwidth]{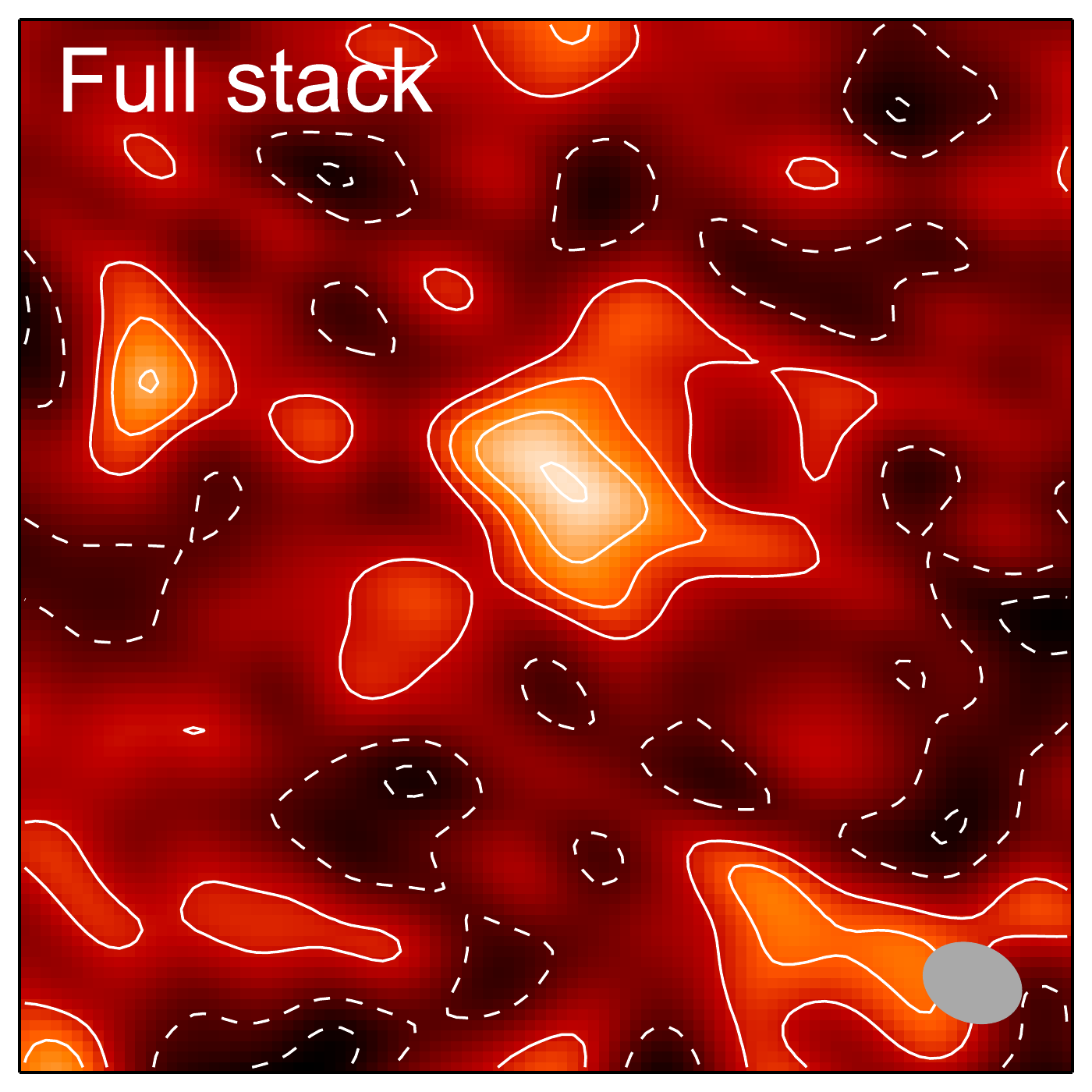}
\includegraphics[width = 0.23\textwidth]{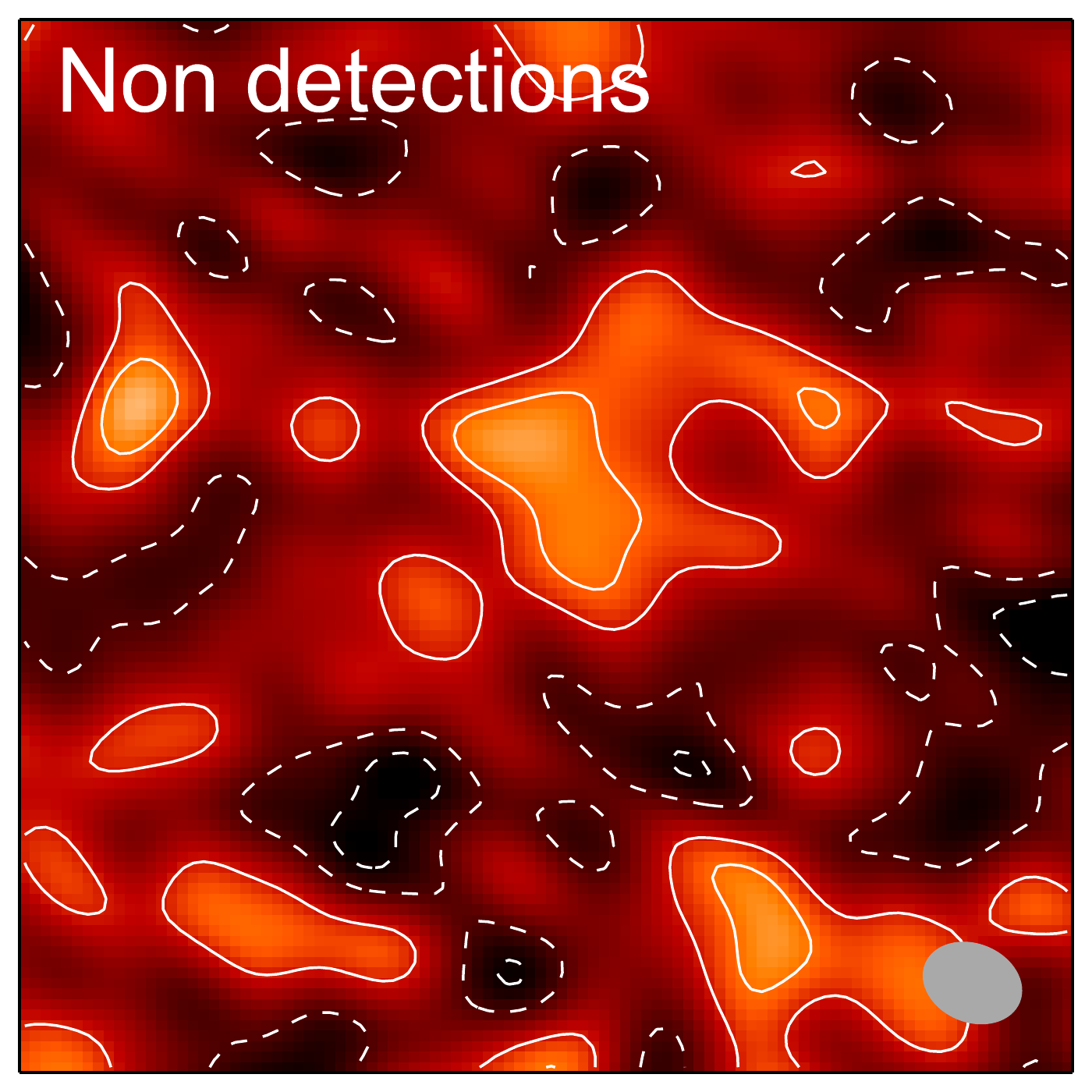}

\caption{The results of stacking the ALMA data for our sample.
The images are 15 arcsec on the side and show the Band 6 data stacked at the centroid position of the LBG in the UltraVISTA $K_s$-band image.
The left-hand plot shows the results of stacking the full sample, whereas in the right-hand stack the detected object ID65666 has been removed.
The colour scale is linear and identical in each plot.
The beam is shown as the filled grey ellipse.
}\label{fig:stack}

\end{figure}

We stacked the ALMA Band 6 data to determine the average FIR luminosity for our sample.
The stacks are shown in Fig.~\ref{fig:stack}.
The data was stacked using the UltraVISTA $K_{s}$-band centroid of each LBG.
The $K_s$-band was used in preference to the~\emph{HST} data as it probes a longer wavelength that is more likely to be representative of the peak in stellar mass, rather than tracing only recent star-formation.
In reality, these coordinates are within $0.1\,$arcsec of the~\emph{HST}/WFC3 coordinates, and hence this makes no difference to our conclusions.
We produced two stacks, one with all six galaxies included, and one where our detected galaxy ID65666 was excluded.
We find a detection at the $4\sigma$ level in the full stack ($f_{\rm 1.3mm} = 46 \pm 11\,\mu{\rm Jy}/{\rm beam}$), and a more tentative detection with a peak at $2.9\sigma$ ($f_{\rm 1.3mm} = 35 \pm 12\,\mu{\rm Jy}/{\rm beam}$) in the stack of the non-detections.
Using the Gaussian fit results in a higher measured flux, indicating that the stacked flux is extended at the resolution of the ALMA data.
This can also be seen visually in the stacks, and could be a consequence of offsets between the FIR and UV emission (Section~\ref{sect:offset}), and/or the FIR emission being resolved in the ALMA data (as we find for ID65666).
We therefore determine the average FIR luminosity of the stacks using the flux determined from the Gaussian fit, and present these results in Table~\ref{table:properties}.
Reassuringly, the results for the full stack and the stack of the non-detections are very similar (although the uncertainty in the derived FIR for the non-detections stack is large), indicating that the detected galaxy does not dominate the stacked flux.
The average FIR luminosity that we derive, $L_{\rm FIR} \simeq 2 \times 10^{11}\,{\rm L}_{\odot}$, is significantly brighter than the majority of the~\citet{Capak2015} sample, which were measured to have $L_{\rm FIR} < 2 \times 10^{10}\,{\rm L}_{\odot}$.
As the UV luminosities for our sample and this $z =5$--$6$ sample are comparable, we therefore derive substantially higher IRX values than found by~\citet{Capak2015}.
We discuss this further in reference to the IRX-$\beta$ relation in Section~\ref{sect:irx}.

\subsection{Serendipitous detections and astrometric accuracy}\label{sect:seren}

\begin{figure}

\includegraphics[width = 0.23\textwidth]{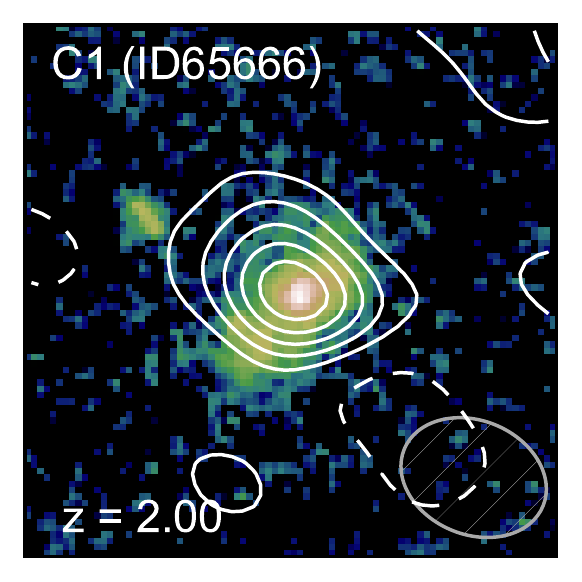}
\includegraphics[width = 0.23\textwidth]{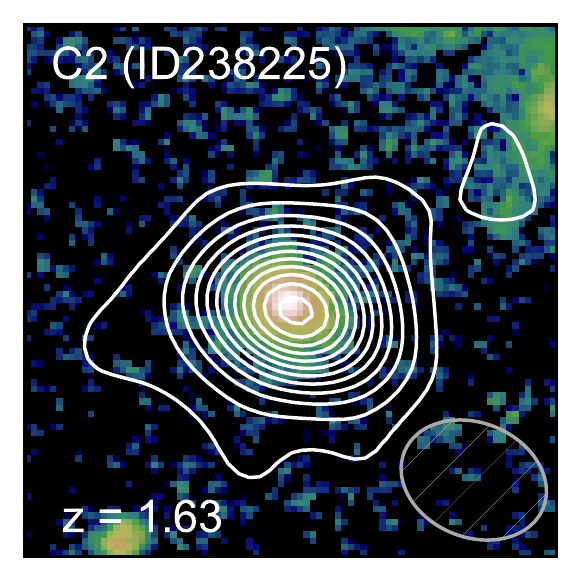}\\
\includegraphics[width = 0.23\textwidth]{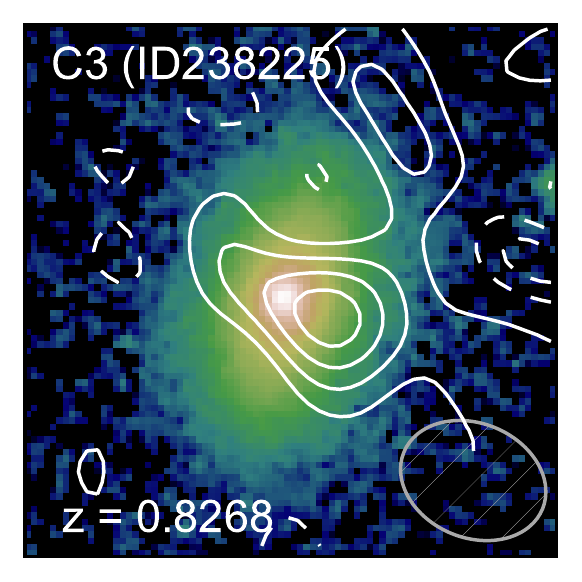}
\includegraphics[width = 0.23\textwidth]{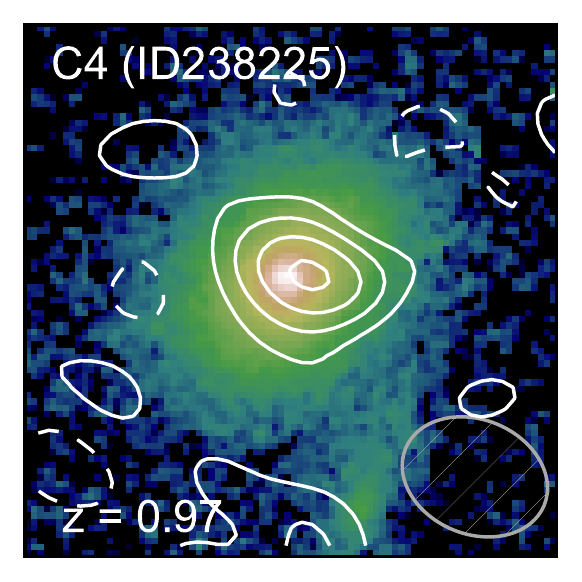}\\

\caption{The four serendipitous ALMA sources detected in the pointings for primary target ID65666 (one object) and ID238225 (three objects).
The best-fitting photometric redshift, or spectroscopic redshift for object C3, is shown in the lower left of each stamp.
Each image is $5\,$arcsec on a side, with the greyscale of the background~\emph{HST}/WFC3 imaging scaled linearly as in Fig.~\ref{fig:stamps}.
For the two bright sources (upper row), the ALMA contours are shown in intervals of $2\sigma$.
}\label{fig:seren}
\end{figure}

The primary beam of the ALMA images has a FWHM of 26 arcsec, and since sensitivity falls off gradually with radial distance from the centre, this allows background sources to be detected serendipitously.
From inspecting the ALMA data we visually identified two bright sources likely corresponding to other galaxies within the field-of-view of the images for primary targets ID65666 and ID238225.
In addition to a visual inspection, we blindly extracted sources from the images following a similar procedure to that of~\citet{Dunlop2017}.
We first created a signal-to-noise ratio (SNR) map to use as the detection image: this is given by the cleaned image (before primary beam correction) divided by the measured RMS in that image.  
Sources were identified by finding the peak SNR pixel and then fitting an elliptical Gaussian profile to this position in the primary beam-corrected image. 
After fitting and subtracting the model from the image, the process was repeated if the residual SNR map contained a peak $>3.5$.
The result was 17 potential sources over the six pointings.
As described in~\citet{Dunlop2017} however, the noise properties of the ALMA data is such that at this significance level we expect more than half of these sources to be spurious.
We therefore cross-matched the ALMA sources with sources detected at $>5\sigma$ significance in the~\emph{HST}/WFC3 $JH_{140}$ image.
The result is a sample of five robust sources from the maps, one of which is the primary target ID65666 at $z = 7.17^{+0.09}_{-0.06}$.
The properties of these sources are detailed in Table~\ref{table:seren}.
The ALMA Band 6 data and the~\emph{HST}/WFC3 imaging for the four additional sources are shown in Fig.~\ref{fig:seren}.
One of the serendipitous detections has been spectroscopically confirmed at $z = 0.8268$ as part of the zCOSMOS program~\citep{Lilly2006}.
This source is also detected in the~\emph{Chandra} X-ray catalogue (`lid\_38';~\citealp{Civano2016}).
For the other sources we quote the photometric redshifts from the COSMOS2015 catalogue~\citep{Laigle2016}.
Three of the sources are detected in the COSMOS Very Large Array (VLA) $3\,{\rm GHz}$ data.

\begin{table*}
\caption{The serendipitous detections found within our ALMA Band 6 dataset.  
The four sources were detected blindly from the data following the procedure outlined in Section~\ref{sect:seren}.
Column 1 shows the primary target ID of the ALMA pointing in which the source was found.
Column 2 is the serendipitous source ID, which is followed by the central coordinates determined from the ALMA data.
The measured flux in the ALMA Band 6 imaging is shown in column 5, followed by an estimate of the $JH_{140}$ magnitude obtained using {\sc MAG\_AUTO} from {\sc Sextractor} in column 6.
The offset in the R.A. and Dec. directions corresponding to the ALMA -~\emph{HST} coordinates are shown in columns 7 and 8.
In columns 9 we show the photometric or spectroscopic redshift, and finally in column 10 we present the VLA and Chandra IDs if available.
}\label{table:seren}
\begin{tabular}{l c c c c c c c c l}
\hline

PID & ID & R.A.(J2000) & Dec.(J2000) & $f_{\rm 1.3mm}$ & $JH_{140}$ & $\Delta_{\rm R.A.}$ & $\Delta_{\rm Dec.}$ & z & Alternative IDs \\
& & & & ($\mu{\rm Jy}$) & (mag) & & & & \\
\hline
65666 & C1 & 10:01:39.75 & +01:54:56.00&$734\pm58$& 22.8 & -0.02 & 0.16 & $2.00_{-0.06}^{+0.13}$& COSMOSVLADP\_J100139.76+015455.8\\[1ex]
238225 & C2 & 10:01:51.71 & +02:25:46.26&$1044\pm44$& 22.9 & -0.12 & -0.04 & $1.63_{-0.06}^{+0.11}$& COSMOSVLADP\_J100151.72+022546.3\\[1ex]
238225 & C3 & 10:01:51.42 & +02:25:31.95&$476\pm100$& 20.1 & -0.45 & -0.19 & $0.8268$& COSMOSVLADP\_J100151.44+022532.1,lid\_38\\[1ex]
238225 & C4 & 10:01:51.59 & +02:25:50.37&$272\pm57$& 20.3 & -0.18 & 0.03 & $0.97_{-0.01}^{+0.01}$& \\
\hline
\end{tabular}
\end{table*}

The detection of these sources, and in particular an object in the same primary beam as our detected high-redshift galaxy ID65666, provides an additional test of the astrometric accuracy.
The astrometry of the~\emph{HST}/WFC3 data was tied to that of the UltraVISTA DR3 images\footnote{\url{https://www.eso.org/sci/observing/phase3/data\_releases/uvista\_dr3.pdf}}, which in-turn is matched to 2MASS~\citep{Magnier2016}.
The expected astrometric accuracy of the two datasets is of order $\sim 0.1\,{\rm arcsec}$, as this is the median offset found when matching the UltraVISTA (and hence~\emph{HST}/WFC3) data to external catalogues.
Such a shift represents $< 1\,{\rm pixel}$ in the ground-based imaging.
We checked the astrometric accuracy of the UltraVISTA DR3 data by matching the full catalogue (covering $1.5\,{\rm deg}^2$) to the AllWISE and COSMOS VLA catalogues, finding excellent agreement and no evidence for a systematic offset larger than $0.05\,{\rm arcsec}$.
In Table~\ref{table:seren} we present the measured astrometric offsets found between the $1.3\,{\rm mm}$ emission and the $JH_{140}$ centroid for the ALMA detected serendipitous sources.
The $JH_{140}$ centroid was determined by {\sc SExtractor}, which calculates the barycentre or first order moment of the light profile~\citep{Bertin1996}.
For the two faintest serendipitous sources (C3 and C4) we find offsets between the ALMA profile and the $JH_{140}$ centroid.
As is evident from the images shown in Fig.~\ref{fig:seren} however, these two sources are particularly faint in the ALMA data ($\sim 5\sigma$) and have an extended~\emph{HST}/WFC3 profile.
C3 and C4 are also close to the edge of the primary beam, and inspection of the ACS/$I_{814}$ imaging available shows no evidence for colour gradients or obscured components at the observed ALMA position.
These factors making the astrometric comparison between the ALMA and~\emph{HST}/WFC3 data more uncertain for these galaxies.
We therefore focus on the two sources, C1 and C2, when discussing any astrometric systematics.
These galaxies are both detected at high signal-to-noise by ALMA and are relatively compact in the~\emph{HST} data.
Visually, these two ALMA bright objects show a good agreement between the peak surface brightness in the~\emph{HST}/WFC3 data and the peak in the $1.3\,{\rm mm}$ profiles.
The measured offsets confirm this, with the largest offset being $\Delta_{\rm Dec.} = 0.16\,{\rm arcsec}$ for C1 (ID65666).
Closer inspection reveals that the $JH_{140}$ profile of this object is asymmetric, which results in the {\sc SExtractor} centroid being weighted to the South-East of the galaxy (Fig.~\ref{fig:seren}).
Note that because these serendipitous sources are at $z \lesssim 2$, the $JH_{140}$ imaging probes the rest-frame wavelengths at $>4000\,$\AA~i.e. the rest-frame optical.
We therefore expect the morphology of these sources to be more representative of the underlying stellar mass structure (rather than transient UV star-forming clumps), which traces the most dust obscured regions and hence the location of the expected FIR emission (e.g.~\citealp{Wuyts2012}).
We therefore find no evidence for any significant systematic astrometric offset between the ALMA data and near-infrared data provided by~\emph{HST}/WFC3 and UltraVISTA DR3.
This lends weight to the $0.6\,$arcsec offset observed for the $z \simeq 7$ galaxy ID65666 being a physical separation, rather than an astrometric systematic (see the discussion in Section~\ref{sect:offset}).

\section{Infrared excess-\boldmath$\beta$ relation} \label{sect:irx}

\begin{figure}

\includegraphics[width = 0.49\textwidth, trim = 0 0 0 0.2cm]{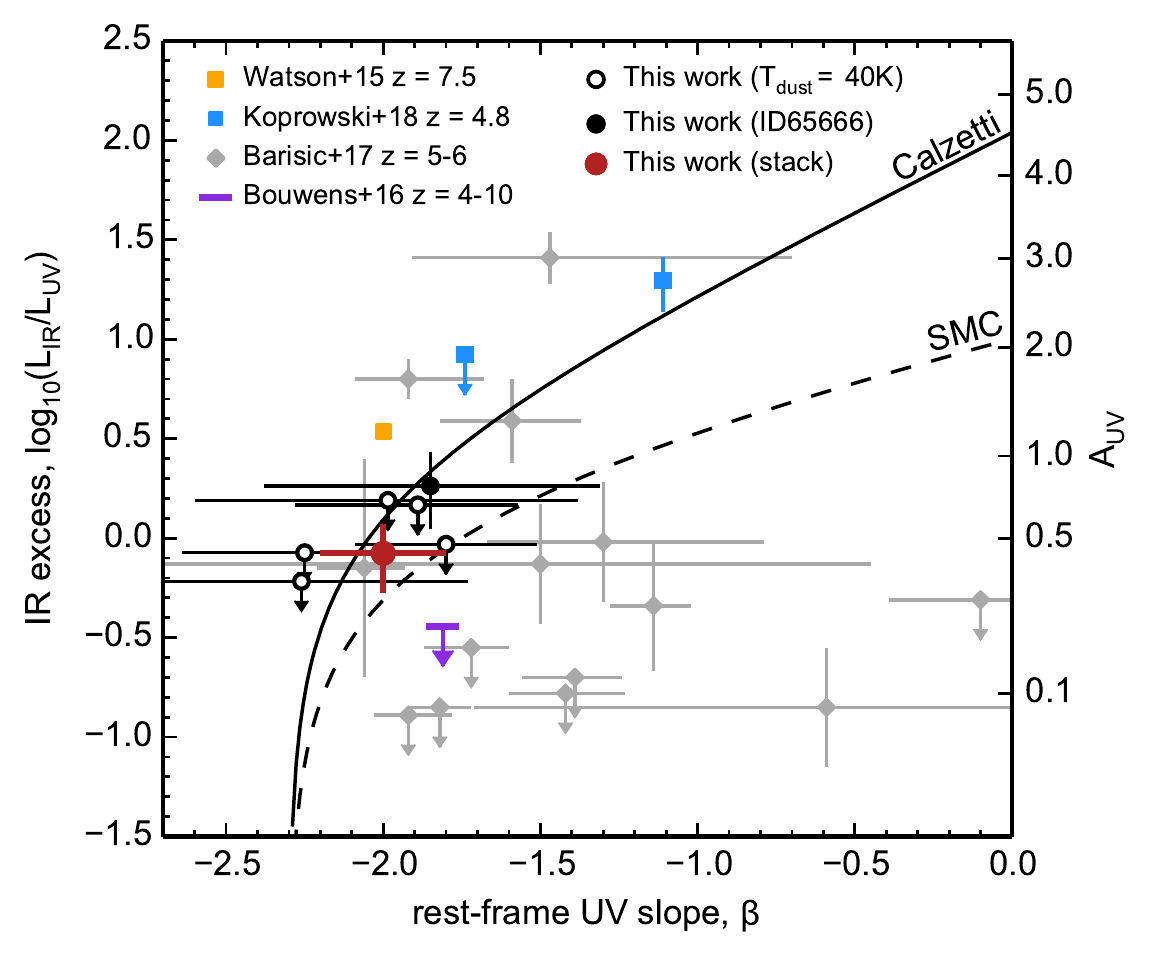}\\
\includegraphics[width = 0.49\textwidth, trim = 0 0 0 0.2cm]{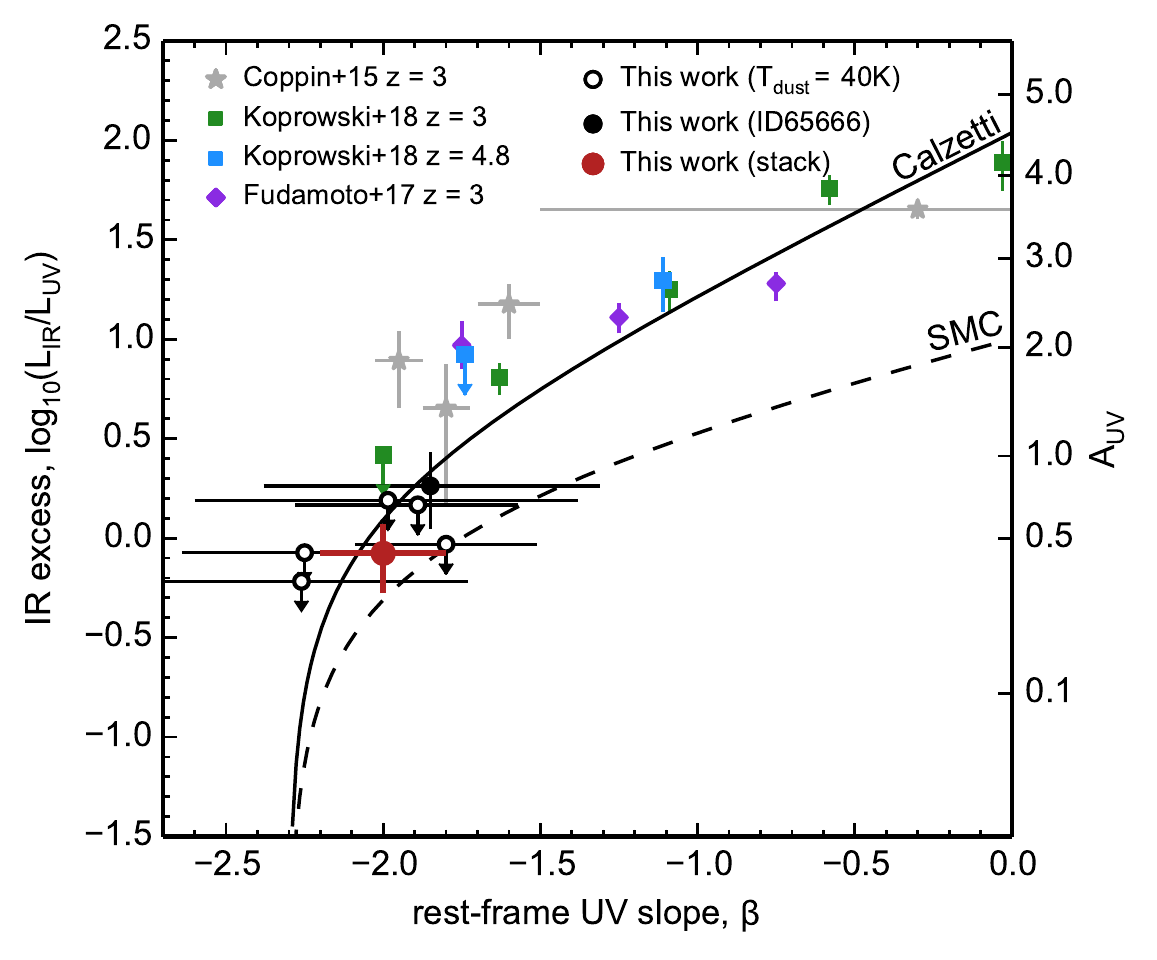}\\
\includegraphics[width = 0.49\textwidth, trim = 0 0 0 0.2cm]{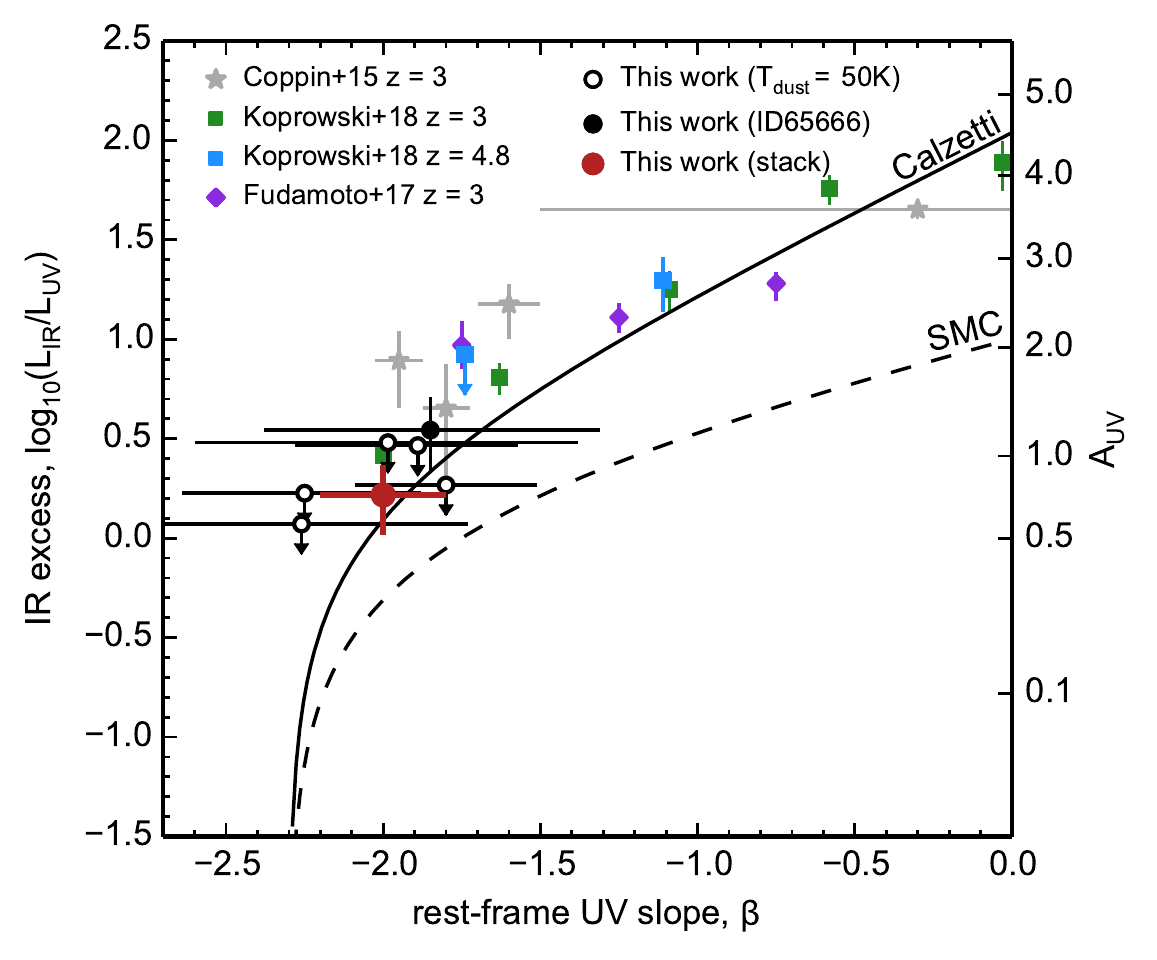}

\caption{The results of this work on the IRX-$\beta$ relation.
We show the detected galaxy ID65666 (filled circle), the non-detected galaxies (open circles) and the result of stacking the full sample (red circle).
Previous studies at $z > 5$ are shown in the upper plot, whereas in the lower plots we compare to studies of $z \simeq 3$--$5$ LBGs.
The final plot shows our results assuming a higher dust temperature of $T = 50\,{\rm K}$.
The IRX-$\beta$ relations are shown assuming the Calzetti (solid line) and SMC dust laws (dashed line).
}\label{fig:irx}

\end{figure}

In order to compare our results to previous studies of the IRX-$\beta$ relation, we calculate the $L_{\rm FIR}$ for our sample by fitting a modified blackbody assuming a dust temperature of T = $40\,{\rm K}$ (see Section~\ref{sect:results}).
The monochromatic rest-frame UV luminosity is calculated at $\lambda = 1600$\AA~as $\nu\,L_{\nu}$, using the measured $M_{\rm UV}$ derived from the $JH_{140}$ data~\citep{Bowler2016}.
The rest-frame UV slope was measured by fitting a power law to the photometry between $\lambda_{\rm rest} = 1270$--$2580$\AA~\citep{Calzetti1994}.
This corresponds to the $Y$, $J$ and $H$ bands at $z \simeq 7$.
For the three highest-redshift galaxies we study, there is also the potential contamination of the $Y$-band by the Lyman-$\alpha$ line (if present) or the Lyman-break.
This is apparent in Fig.~\ref{fig:sed}, where the HSC $y$-band, which does not extend as far red-ward as the VISTA $Y$-band, can be seen to drop in the three galaxies which have $z_{\rm phot} \gtrsim 6.9$.
Hence we additionally exclude the $Y$ band in the fitting for ID65666, ID238225 and ID304416.
For the stack, we plot the mean $\beta$ slope of the included galaxies, with the error given by the standard error on the mean.
As the results from the full stack and the stack of the non-detections are similar, we show the full stacked results on the IRX-$\beta$ relation.
We calculated the stacked IRX values in two ways; 1) we took the average $L_{\rm FIR}$ determined from the stack, and divided this by the average $L_{\rm UV}$ of the sample and 2) we normalised the ALMA images according to the IRX of each galaxy and then stacked these normalised images to determine the stacked IRX.
The results of these two methods were consistent to within $\Delta {\rm log}({\rm IRX}) = 0.01$, and hence we choose to present the results of method (1) here.
We show the results from this study on the IRX-$\beta$ diagram in Fig.~\ref{fig:irx}.
On this plot we also present the~\citet{Meurer1999} relation derived at low redshift, assuming both a starburst~\citet{Calzetti2000} attenuation law and a steeper SMC extinction law.
Both IRX-$\beta$ relations are plotted assuming an intrinsic rest-frame slope of $\beta = -2.3$ and following the form presented in~\citet{McLure2017}.

The only other work that has studied a sample of comparably bright ($L_{\rm UV} > 10^{11}\,{\rm L}_{\odot}$; see Fig.~\ref{fig:luvlir}) LBGs at $z >5$ was undertaken by~\citet{Capak2015}, who selected their sample of ten $z = 5$--$6$ galaxies from the ground-based COSMOS field.
In~\citet{Barisic2017}, updated rest-frame UV slopes for the galaxies studied in~\citet{Capak2015} were presented, and hence we compare to this more recent study hereafter.
As is evident in the upper panel of Fig.~\ref{fig:irx}, the results from our sample of bright $z \simeq 7$ LBGs occupy a markedly different region of the IRX-$\beta$ plot than the~\citet{Barisic2017} results.
Our results do not extend to the extremely low IRX values found by~\citet{Barisic2017}, which are $>0.5\,{\rm dex}$ lower than our stack at $z \simeq 7$.
Note that if we use an identical parameterisation of the FIR SED as~\citet{Barisic2017} (see Section~\ref{sect:results}) our derived IRX values would increase by $0.2$ dex, further highlighting this discrepency.
We also do not find the large scatter to high IRX values, at a given $\beta$, found at $z \simeq 5$.
In comparison to the spread of rest-frame UV slopes found by~\citet{Barisic2017}, we find a tighter range in $\beta$ in our sample, although the large errors on $\beta$ in both studies means the majority of the galaxies have formally consistent rest-frame UV slopes.
The six LBGs in this study show blue rest-frame UV slopes around $\beta \simeq -2$.
This could potentially be due to selection effects (e.g.~\citealp{Rogers2013, Dunlop2013}), where redder galaxies are more likely to be classified as contaminant populations (such as dusty $z \simeq 2$ galaxies) and hence be removed from the sample.
These objects are very bright however, being detected up to $2\,{\rm mag}$ brighter than the limiting magnitudes of the UltraVISTA DR2 data (used for the initial selection), leading to a high completeness as a function of $\beta$ slope~\citep{Bowler2014}.
Given the short cosmic time between $z = 7$ and $z = 5$ ($\simeq 400\,$Myr), it is somewhat surprising that there is such a large difference in these results.
The~\citet{Capak2015} sample contains both narrow-band selected Lyman-$\alpha$ emitters (LAEs) as well as LBGs, which could contribute to the scatter in IRX (e.g.~\citealp{Schaerer2015}), however~\citet{Barisic2017} show that both the LAEs and LBGs with weak or no Lyman-$\alpha$ emission show similarly low IRX values.
It is possible that using ground-based photometry to measure $\beta$, as opposed to higher resolution data from~\emph{HST}/WFC3, could bias our measurements to bluer values.
This is because we measure $\beta$ at the peak position of the rest-frame UV emission, which is expected to be the least obscured and hence have the bluest colour (e.g. we measure a `light-weighted' colour).
In the extreme, a dusty star-forming region of a galaxy can become completely obscured in the rest-frame UV, an effect we discuss further in Section~\ref{sect:offset}.
We cannot measure a resolved $\beta$ from our current dataset, as we only have one~\emph{HST}/WFC3 filter ($JH_{140}$).
The comparison of $\beta$ values derived from the $z \simeq 5$--$6$ sample of~\citet{Capak2015} however, suggest that this effect is not the dominant systematic error on $\beta$, as the ground-based measurements in~\citet{Capak2015} were found to be redder on the whole, not bluer, than the resolved measurements from~\emph{HST}/WFC3 colours presented in~\citet{Barisic2017}.
Note that taking into account the potential offset between the rest-frame UV and FIR emission in ID65666 (Section~\ref{sect:offset}) does introduce a larger range in the derived IRX for this object.
Ultimately larger samples of homogeneously selected LBGs with deep ALMA~and resolved rest-frame UV imaging will be required to understand these differences.

In Fig.~\ref{fig:irx} we also compare to the stacked results of~\citet{Bouwens2016b}.
Using the ASPECS data in the~\emph{Hubble} Ultra Deep Field,~\citet{Bouwens2016b} found upper limits on the IRX of ${\rm log}({\rm IRX}) \lesssim -0.4$ ($2\sigma$; $T = 35\,$K) for galaxies at $z = 4$--$10$.
This limit is clearly in tension with the average IRX we find from our stack (and more so with our detection for galaxy ID65666).
\citet{McLure2017} have shown that stacking analyses which use bins of $\beta$ rather than stellar mass are prone to biases that act to artificially lower the IRX in a given $\beta$-bin.
We therefore only show the full stacked result from~\citet{Bouwens2016b}, rather than the binned results.
This point includes galaxies with a wide range in measured rest-frame UV slope, from $ \beta \simeq -4$ to $\beta > -1.25$, and hence the single point should be treated with caution.
Furthermore, the majority of the galaxies probed in the ASPECS data were fainter in the rest-frame UV than those studied here, with apparent magnitudes peaking between $m_{\rm AB} = 28$--$30$ (as opposed to $m_{\rm AB} \simeq 24$--$25$ for our sample).
Nevertheless, the results of this study appear to disagree strongly with the limits obtained from the ASPECS data.
Instead (as we discuss below) our results are consistent with the Calzetti-like attenuation law found in lower-redshift galaxies.
Similarly, the recent stacking results of~\citet{Koprowski2018} at $z = 4.8$ are in good agreement with the Calzetti-law, with their study finding no evidence for any evolution in the normalisation of the IRX-$\beta$ relation from $z =3$--$5$.
Finally, our stacked result is significantly lower than the IRX derived for the lensed galaxy `A1689-zD1'~\citep{Watson2015} at $z = 7.5$.
While the FIR luminosity we determine for our stack is higher than the~\citet{Watson2015} and~\citet{Laporte2017} objects at $z > 7$, because our sample is more luminous in the UV, the IRX values we derive are lower than these studies.
Our results therefore suggest that these two very high-redshift dust detections are not representative of the $z \gtrsim 6$ population as a whole.

Given the small samples sizes and physical differences identified in the previous results at $z > 5$, we also compare our results to previous studies of IRX-$\beta$ in $z \simeq 3$--$5$ LBGs.
In the middle and bottom panels of Fig.~\ref{fig:irx} we show the results of~\citet{Fudamoto2017},~\citet{Coppin2015} and~\citet{Koprowski2018}, which were all derived from LBGs selected over degree-scale ground-based fields.
As such, these $z \sim 3$ samples have a similar UV luminosity and number density to our sample.
The results of our study at $z \simeq 7$ are consistent with these previous determinations of IRX-$\beta$, overlapping with the blue-end of these data points (e.g. from~\citealp{Koprowski2018}) at the point where the IRX-$\beta$ relation is predicted to rapidly drop-off.
Recent results have indicated that the results at $z \simeq 3$ support a Calzetti-like attenuation curve, with little evidence for a SMC-like extinction law once biases have been corrected for~\citep{McLure2017, Koprowski2018}.
In common with these findings, our derived IRX-$\beta$ points for the individual $z \simeq 7$ LBGs are consistent with a Calzetti-like dust attenuation curve, shown as the solid line in Fig.~\ref{fig:irx}.
While there is a large error on the $\beta$ measurement, the results for galaxy ID65666 are also as expected from the Calzetti-law, showing an FIR luminosity that is a factor of two brighter than that expected if an SMC-like extinction curve was in effect.
With the assumption of a dust temperature of T $=40\,$K our stacked result is consistent with both a Calzetti and SMC-like dust attenuation/extinction law, with a slightly better agreement with the Calzetti relation.
While it has yet to be directly observed in $z > 5$ LBGs, several observational works have suggested that the dust temperature at these redshifts could be substantially higher (e.g.~\citealp{Bethermin2014, Schreiber2017}).
Indeed, some theoretical models/simulations indicate the dust temperature could be as high as $\simeq 60\,{\rm K}$ at $z = 7$~\citep{Narayanan2018, Imara2018}, with potential variations in $T$ between individual galaxies in the same sample~\citep{Pavesi2016}.
We therefore also present the derived IRX-$\beta$ results for our sample assuming a higher dust temperature of T$ = 50$K in the bottom panel of Fig.~\ref{fig:irx}.
The result is a vertical shift of $0.3$ dex due to the derived $L_{\rm FIR}$ being double that obtained assuming our fiducial dust temperature.
Clearly in the case of an evolving dust temperature, our results are in excellent agreement with the $z \simeq 3$ Calzetti-like relation.
In this case our stacked results exclude an SMC-like extinction law at the $\gtrsim 2\sigma$ level.
In conclusion, we find no compelling evidence for extremely low IRX values at $z > 5$ as has previously been claimed by~\citet{Capak2015} and~\citet{Bouwens2016b} from our sample of UV bright LBGs at $z \simeq 7$.
Instead, particularly if the dust temperature is evolving with redshift, our results are fully consistent with a Calzettti-like dust attenuation curve as found in local star-burst galaxies.

\section{Discussion}\label{sect:diss}

\subsection{A physical offset between the UV and FIR emission}\label{sect:offset}

When comparing the~\emph{HST}/WFC3 $JH_{140}$ band image to the ALMA $1.3\,{\rm mm}$ detection in object ID65666, we find an offset of $0.6\,$ arcsec predominantly in the North-South direction ($\Delta_{\rm R.A.} = 0.57\,$arcsec, $\Delta_{\rm Dec.} = 0.17\,$arcsec).
Offsets between the dust continuum and rest-UV emission have been identified in several high-redshift LBGs~\citep{Laporte2017, Faisst2017, Koprowski2016}, as have offsets between the {\sc [CII]} FIR line and the rest-UV continuum (e.g.~\citealp{Carniani2017, Maiolino2015, Willott2015, Capak2015}).
Some of these offsets have been attributed to astrometric systematics (e.g.~\citealp{Dunlop2017}), however there is a growing consensus that FIR emission lines in particular are physically offset at $ z \simeq 7$ (see~\citealp{Carniani2017a}).
The offset we observe corresponds to a physical separation of $\simeq 3\,{\rm kpc}$.
This separation is comparable in magnitude to what has been found previously in high-redshift LBGs (e.g. in HZ4 and HZ10; \citealp{Faisst2017}).
The physical nature of this offset is strengthened by the presence of a bright ALMA source in the same primary beam as ID65666 (Fig.~\ref{fig:seren}), which shows no significant astrometric offset (see Section~\ref{sect:seren}).
The expected positional accuracy in the ALMA data is given as $0.6 \times {\rm FWHM}/{\rm SNR}$~\citep{Ivison2007}, which corresponds to $\sigma = 0.17$ along the major axis and $\sigma = 0.13$ along the minor axis of the beam.
We therefore find that the offset has a significance of $\simeq 4\sigma$ in the absence of systematic errors (which our checks have demonstrated are minimal).
Note that in~\citet{Carniani2017a} they shift the near-infrared data such that serendipitous sources are perfectly aligned with the ALMA data.
In this case, the observed astrometric offset observed for ID65555 would be $0.45\,$arcsec, which is still a $\sim 3\sigma$ separation.

\begin{figure}

\includegraphics[width = 0.49\textwidth]{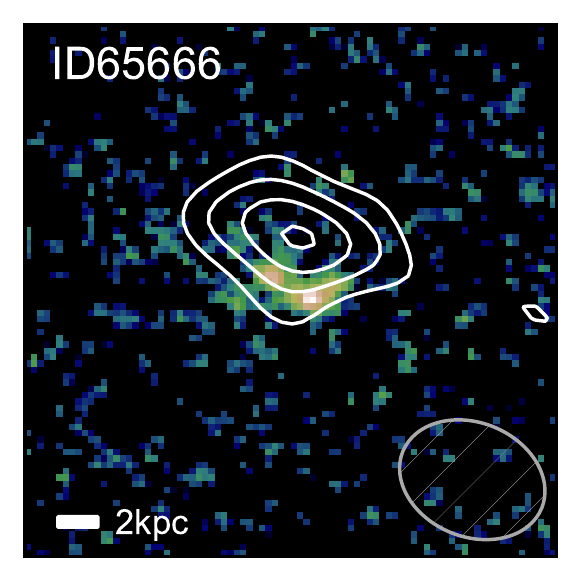}\\
\caption{The FIR detected galaxy, ID65666, at $z = 7.17^{+0.09}_{-0.06}$.
The background image shows the~\emph{HST}/WFC3 $JH_{140}$ data probing the rest-frame UV emission.
The ALMA Band 6 data probing the rest-frame FIR dust continuum is shown as the contours, which are displayed upwards and including $2\sigma$ significance.
An offset of 0.6 arcsec is seen between the datasets, corresponding to a physical separation of $3\,{\rm kpc}$ (see Sections~\ref{sect:seren} and~\ref{sect:offset} for discussion).
The image is $5\,$arcsec on the side, and the beam is shown as the grey ellipse in the lower right-hand corner.
}\label{fig:zoom}
\end{figure}

As shown in the enlarged image in Fig.~\ref{fig:zoom}, the rest-frame UV emission for ID65666 shows a clumpy, extended morphology.
In comparison to the ALMA contours, the observed rest-frame UV emission appears to wrap-around the central obscured part, as is seen in dusty star-forming galaxies and sub-mm galaxies at low redshift (e.g.~\citealp{Hodge2015, Chen2017}).
In the prediction of a relationship between the IRX and the rest-frame UV slope, there is the assumption that the stars and dust are well mixed, leading to the coupling of any observed reddening in the UV to the FIR emission detected (e.g.~\citealp{Charlot2000}).
If instead, the galaxy consists of regions of significantly different obscuration, then the relationship will break down for the galaxy as a whole.
At the central position of the $1.3\,{\rm mm}$ detection for ID65666 we find no evidence for rest-frame UV emission in the~\emph{HST}/WFC3 data.
This ``UV-dark'' part of this LBG therefore has a high IRX (and potentially very red $\beta$), which is significantly different to the value derived for the full galaxy. 
Taking the $2\sigma$ upper limit of the $JH_{140}$ data as $m_{\rm AB} = 27.9$ (equivalent to $M_{\rm UV} \lesssim -19$), we estimate that the IRX of this obscured region is IRX $\gtrsim 1.5$ corresponding to an $A_{\rm UV} \simeq 3\,$mag.
Conversely, for the UV-visible part of the LBG found within the~\emph{HST} data, the IRX value is lower than that calculated for the full object if the dust emission is physically decoupled.
If we assume that there is no $1.3\,{\rm mm}$ emission at the $2\sigma$ level from the UV-bright part of the LBG, this results in an upper limit on the IRX $< -0.35\,(2\sigma)$ which is still consistent with what would be expected the IRX-$\beta$ relation found at lower redshift~(given the errors on $\beta$).
Thus within ID65666, we find a considerable range in the measured IRX across the galaxy (see also~\citealp{Koprowski2016} and the discussion in~\citealp{Faisst2017}), and evidence that the observed rest-UV emission is spatially distinct from the site of strong dust emission.
Such a decoupling of the galaxy colour and FIR emission has been observed in infrared-selected galaxies, where above $L_{\rm FIR} > 10^{11}\,L_{\odot}$ galaxies appear bluer than expected from the IRX-$\beta$ relation~\citep{Casey2014a}.

In our stacking analysis, we find evidence that the FIR flux is extended when stacking is performed based on the rest-frame UV position.
If offsets between the rest-UV and rest-FIR emission like that observed for ID65666 are common in our sample, then this would provide a natural explanation for the extended flux observed in the stack (Fig.~\ref{fig:stack}).
If spatial offsets of the order of $\sim 0.5\,$arcsec are common within the LBG population (as they are in the SMG population;~\citealp{Chen2017, Riechers2014}), this could result in a reduced sensitivity to flux when stacking based on the rest-frame UV position.
To test this hypothesis we performed a simple simulation of stacking artificial ALMA sources with spatial offsets in random directions. 
We assumed a Gaussian profile with a FWHM of $1$ arcsec, to approximate the ALMA beam in our observations.
In the case of a constant $0.5\,$arcsec offset, we find that the S/N ratio was approximately half that expected if the rest-frame UV and FIR centroid were the same (assuming naturally weighted images).
In the case of a distribution of offsets, chosen randomly between $0$--$0.5\,$arcsec, the stacked flux was only marginally lower than with no assumed offset.
Clearly the result depends sensitively on the resolution of the FIR datasets (such that SCUBA-2 stacking for example, would be unaffected), on the magnitude of the offset and on the method used to measure the flux.
We therefore conclude that such an effect is unlikely to be important for the fainter LBG population at high-redshift, which tend to show compact and smooth profiles~\citep{Curtis-Lake2016}, but could be significant for bright galaxies where the observed dust and UV emission are extended on the scale of $\simeq 1\,$ arcsec.

\subsection{The SFR and ${\mathbf M_{\star}}$ of bright LBGs at \boldmath $z = 7$}

We calculated the FIR star-formation rate using the calibration from~\citet{Kennicutt2012}, correcting to a~\citet{Chabrier2003} initial-mass function (IMF).
The star-formation rate from the observed (uncorrected) rest-frame UV emission was estimated using the~\citet{Madau1998} prescription corrected to a~\citet{Chabrier2003} IMF.
The UV SFR was found to be in the range $17$--$34$\SFRu for the sample~\citep{Bowler2016}.
For the stack we find that the obscured SFR (SFR$_{\rm FIR} \simeq 30$\SFRu) is comparable to that observed in the rest-frame UV.
Clearly therefore, dust obscured star-formation is significant in bright $z \simeq 7$ LBGs.
Furthermore, if the dust temperature is higher at these redshifts than our fiducial value, then the SFR$_{\rm FIR}$ would be even higher (by a factor of two for T$\,=50\,$K).
In the individually detected galaxy ID65666, the obscured component of the SFR is even larger, with $75$\% of the total SFR being obscured (increasing to $85$\% in the case of T$ \,= 50\,$K).

In comparison to the full sample, our FIR detected object ID65666 appears unremarkable in terms of SFR$_{\rm UV}$ and rest-frame UV slope.
In-fact it has one of the lower SFRs as derived from the observed magnitude in the rest-frame UV.
Recent results at $z \simeq 3$ have highlighted the importance of stellar mass in predicting the dust attenuation, or IRX, of star-forming galaxies~\citep{Dunlop2017, McLure2017, Bourne2017}.
We therefore estimated the stellar masses of the galaxies in our sample, with the expectation that ID65666 would be one of the most massive.
We performed SED fitting to the available optical, near- and mid-infrared data.
Crucially this includes~\emph{Spitzer}/IRAC photometry which probes the rest-frame optical emission at $z \simeq 7$.
At these redshift however, strong rest-frame optical emission lines of H$\alpha$, H$\beta$ and \oiii can contaminate the \chone and \chtwo bands and artificially boost the derived masses.
The SED fits are shown in Fig~\ref{fig:sed} and the derived stellar masses in Table~\ref{table:properties}.
We find evidence for strong rest-frame optical emission lines in our sample~\citep{Bowler2016}.
To provide estimates of the masses, we therefore excluded bands that were potentially contaminated based on the galaxy redshift.
For the lowest-redshift galaxy in our sample, ID279127, potentially both of the~\emph{Spitzer}/IRAC~\chone and \chtwo bands contain strong emission lines.
For this galaxy we fit to the \chtwo-band only, which contains {\sc H}$\alpha$ at $z \lesssim 6.6$, and caution that in this case the stellar mass represents an upper limit.
The stellar masses of our sample are found to be ${\rm log}(M_{\star}/{\rm M}_{\odot)} \simeq 9.5$.
In comparison to the IRX-$M_{\star}$ relation, our sample lies below the relation derived at $z \simeq 2$ by~\citet{McLure2017} as has been found for other high-redshift samples~\citep{Faisst2017, Fudamoto2017}.
The tension would be reduced if the dust temperature is higher at these redshifts, and considering the large errors present in the determination of stellar masses at these redshifts, we cannot make strong conclusions in this study about the presence of an IRX-$M_{\star}$ relation at $z \simeq 7$.

Galaxy ID65666 shows a particularly low mass of ${\rm log}(M_{\star}/{\rm M}_{\odot}) = 9.0^{+0.8}_{-0.2}$ (although the error is large due to the low significance detection in the $[3.6]$-band), and also shows the reddest \chone - \chtwo colour, indicative of particularly strong rest-frame optical emission lines.
Note that the mass we measure is determined from the unobscured part of the LBG, hence could be underestimated.
Using the flux excess in the \chone band compared to the best-fitting continuum model, we estimate that the rest-frame equivalent width of the combined {\sc H}$\beta$ + \oiii lines is $EW_{0} = 1400 \pm 300\,$\AA~for this galaxy (uncorrected for dust attenuation), making it comparable to some of the most extreme nebular emitters known at these redshifts~\citep{RobertsBorsani2015, Bowler2016}.
The properties of ID65666 suggest that it is a dusty starburst, showing an anisotropic distribution of unobscured and obscured star-formation on the scale of several {\rm kpc}.
As such it has a similar morphology and properties (although with weaker dust emission by a factor of 100) as has been observed for high-redshift sub-mm galaxies (e.g.~\citealp{Chen2017, Hodge2015, Casey2014}).

\subsection{Dust in galaxy formation simulations at high redshifts} \label{sect:sims}

In~\citet{Bowler2015} we presented a comparison between the observed rest-frame UV luminosity function at $z = 5$--$7$ and that predicted by a range of cosmological galaxy formation simulations.
One notable trend was that the majority of these simulations over predicted the number density of Lyman-break galaxies bright-ward of $M_{\rm UV} \lesssim -20$.
This excess of galaxies at the bright-end is present in hydrodynamical simulations (e.g. Illustris;~\citealp{Genel2014}, FIRE;~\citealp{Ma2018}, and the First Billion Years model;~\citealp{Cullen2017}), semi-analytic models (e.g. the Munich L-galaxies model;~\citealp{Clay2015}, GALFORM;~\citealp{GonzalezPerez2013} and the~\citealp{Somerville2012} model), as well as the analytic model of~\citet{Cai2014}.
Within these simulations, the observed number densities were only reproduced with the addition of a considerable amount of dust, which acts to suppress the rest-frame UV emission.
For example, in the Munich model presented in~\citet{Clay2015}, galaxies with an observed $M_{\rm UV} < -22$ are predicted to have a UV attenuation of $A_{\rm UV} \simeq 2$ magnitudes.
In the case of a uniform distribution of dust covering the star-forming regions, such an attenuation would result in red rest-frame UV colours of $\beta \gtrsim -1.5$ depending on the dust attenuation law assumed.
At $z \simeq 5$,~\citet{Cullen2017} have shown using the FiBY model, that such an attenuation is consistent with the observed colour-magnitude relationship, with the brightest UV galaxies being attenuated more and showing redder colours.

In the case of well mixed stars and dust, the IRX is directly related to the attenuation in the UV, whereas the relationship between IRX and $\beta$ depends on the dust attenuation law.
It is therefore possible to directly test the expected attenuation using the measurements of the FIR dust continuum for our sample.
We show the $A_{\rm UV}$ on the right-hand $y$-axis of Fig.~\ref{fig:irx}.
The measured FIR emission from our sample implies an average UV attenuation of $A_{\rm UV} \simeq 0.5$--$1.0\,$mag (depending on dust temperature).
Our results therefore show that the presence of dust is significant in suppressing the observed UV luminosity in the brightest galaxies.
While the average magnitude of the attenuation is lower than that predicted in some of the models described above, it is sufficient to have a significant effect on the observed number counts of UV bright LBGs observed at these redshifts.
Moreover, the detection of completely obscured star-formation in galaxy ID65666 in this work, highlights the possibility that some star-forming galaxies may be completely obscured in the rest-frame UV, even at redshifts as high as $z = 7$.
If this is the case, then optical/near-infrared surveys for high-redshift LBGs will form an incomplete census of star-formation.
Our results indicate that a combination of dust obscuring around $50$-$85$\% of the star-formation in bright LBGs, plus the presence of ``UV-dark'' galaxies that are missing from current UV selected samples, can explain the observed shape of the bright-end of the rest-frame UV LF at $z \simeq 7$.

\section{Conclusions}\label{sect:conclusions}

We present new ALMA imaging at $1.3\,{\rm mm}$ (rest frame $\simeq 170\,\mu{\rm m}$) of six bright Lyman-break galaxies at $z \simeq 7$.
The sample was originally selected from $1\,{\rm deg}^2$ of the COSMOS field, and represents some of the most intrinsically UV luminous galaxies at this epoch ($M_{\rm UV} < -22$).
We observe the sample with 10 minute integrations in ALMA Band 6.
The highest redshift galaxy targeted, ID65666 at $z = 7.17_{-0.06}^{+0.09}$, is detected at $5\sigma$ significance at peak emission, corresponding to a FIR luminosity of $L_{\rm FIR} \simeq 3.6 \pm 1.2 \times 10^{11}\,{\rm L}_{\odot}$.
The five remaining galaxies are undetected at the $3\sigma$ level.
The average FIR luminosity of the sample, determined from a stacking analysis, is $L_{\rm FIR} \simeq 2 \times 10^{11}\,{\rm L}_{\odot}$.
Converting this observed $L_{\rm FIR}$ into a star-formation rate indicates that in these galaxies, approximately half of the total SFR is obscured by dust.
In comparison to the IRX-$\beta$ relation, we do not reproduce the extremely low IRX values found at $z > 5$ by several previous studies~\citep{Capak2015, Bouwens2016b}.
Instead, our results are consistent with the predictions of the Calzetti-like dust attenuation law found for star-forming galaxies at lower redshifts.
We find that the presence of dust in the brightest LBGs is significant in shaping the observed UV luminosity function at the bright-end, with galaxies of absolute magnitude $M_{\rm UV} < -22$ being attenuated by $A_{\rm UV} = 0.5$--$1.0$ on average.
Such an attenuation is comparable to that typically applied in galaxy formation simulations.
In the galaxy ID65666 at $z = 7.17_{-0.06}^{+0.09}$, we detect a physical offset between the rest-frame FIR dust continuum probed by ALMA and the rest-frame UV emission probed by~\emph{HST}/WFC3 $JH_{140}$ imaging.
The offset is measured to be 0.6 arcsec, corresponding to a physical separation of $3\,{\rm kpc}$.
The detection of a serendipitous source in the same ALMA pointing rules out any significant astrometric systematic.
The presence of this offset indicates an inhomogeneous distribution of dust within this galaxy, with a high proportion of the total SFR ($>75$\%) being completely obscured in the rest-frame UV.
The existence of this obscured component in galaxy ID65666 illustrates that entire star-forming regions, or potentially even entire galaxies, could be ``UV-dark'' even at $z \simeq 7$.

\section*{Acknowledgements}

This work was supported by the Oxford Hintze Centre for Astrophysical Surveys which is funded through generous support from the Hintze Family Charitable Foundation.
NB acknowledges support from the European Research Council Advanced Investigator Program, COSMICISM (ERC-2012-ADG\_20120216, PI R.J.Ivison).
This paper makes use of the following ALMA data: ADS/JAO.ALMA\#2015.1.00540.S.  
ALMA is a partnership of ESO (representing its member states), NSF (USA) and NINS (Japan), together with NRC (Canada), MOST and ASIAA (Taiwan), and KASI (Republic of Korea), in cooperation with the Republic of Chile. The Joint ALMA Observatory is operated by ESO, AUI/NRAO and NAOJ.
This work is based in part on observations made with the NASA/ESA Hubble Space Telescope, which is operated by the Association of Universities for Research in Astronomy, Inc, under NASA contract NAS5- 26555.
This research made use of {\sc Astropy}, a community-developed core Python package for Astronomy (Astropy Collaboration, 2013).




\bibliographystyle{mnras}
\bibliography{library_abbrv} 

\appendix

\section{SED fitting results}

\begin{figure*}

\includegraphics[width = 0.4\textwidth]{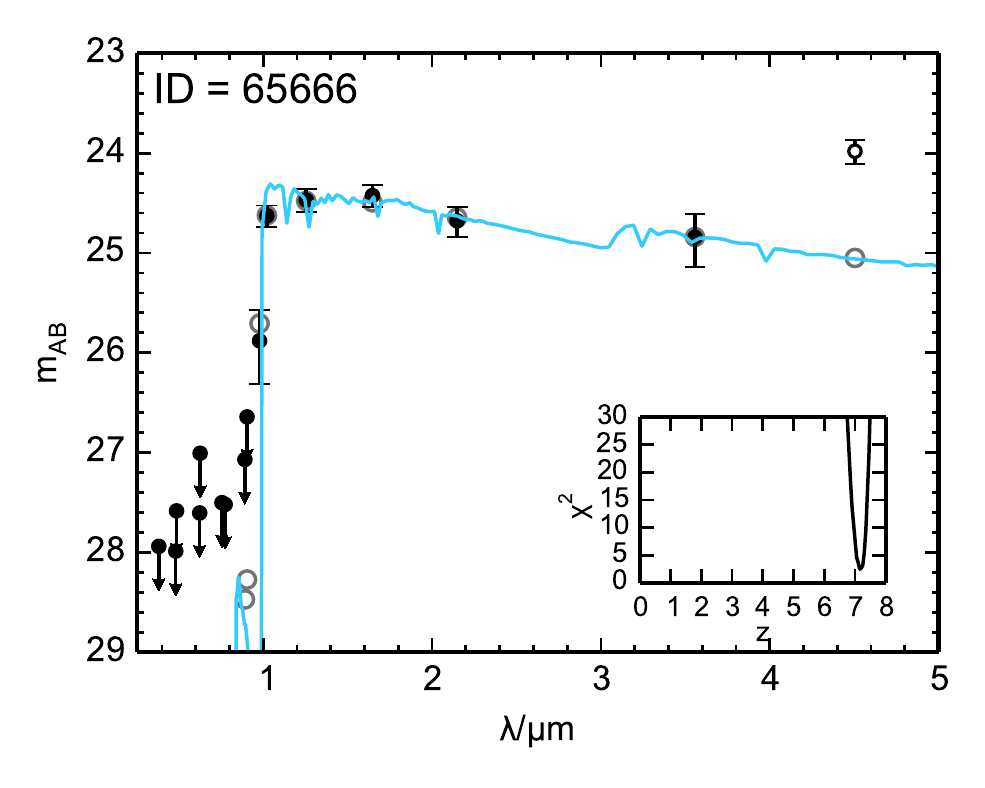}
\includegraphics[width = 0.4\textwidth]{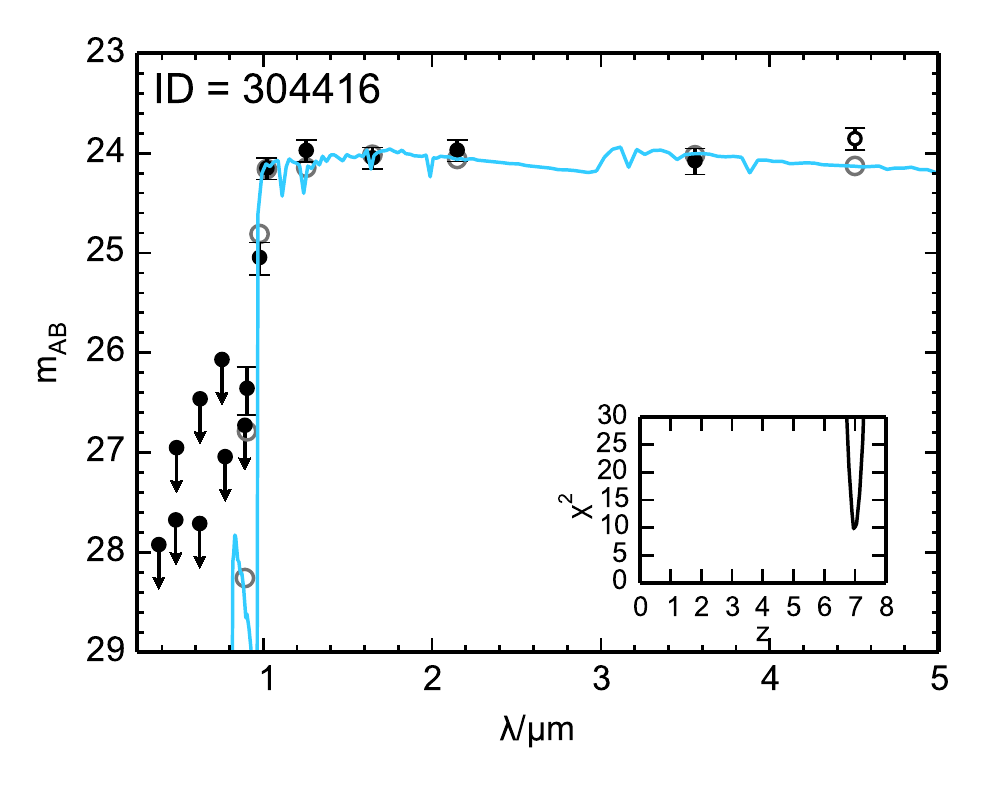}\\
\includegraphics[width = 0.4\textwidth]{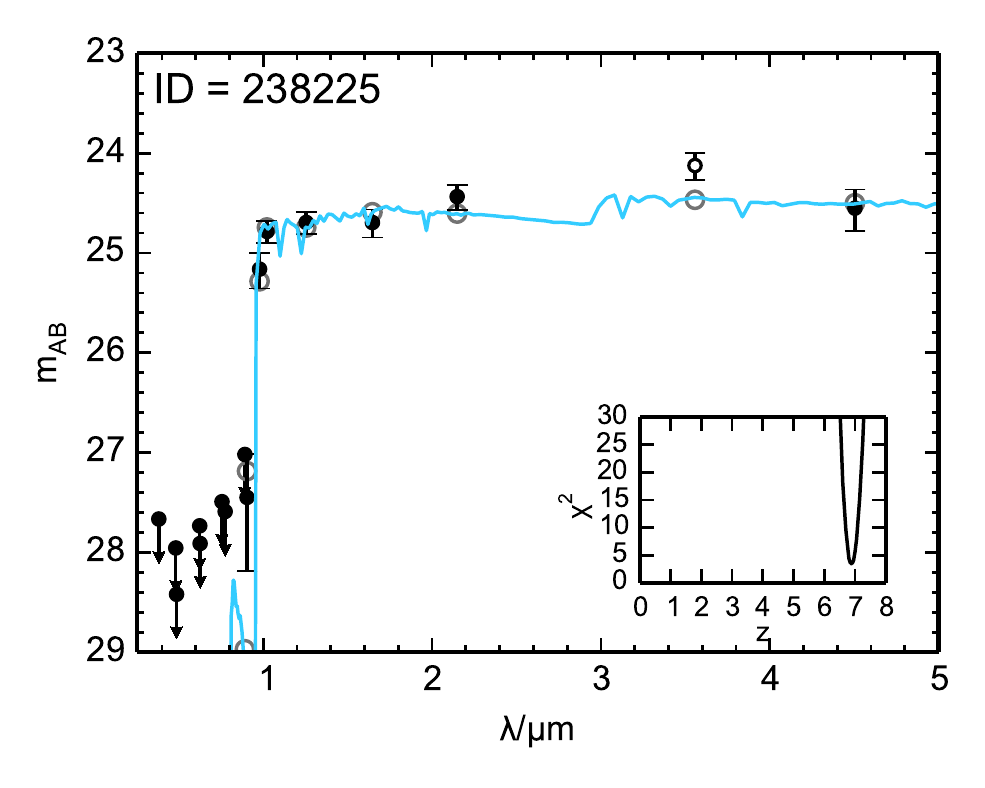}
\includegraphics[width = 0.4\textwidth]{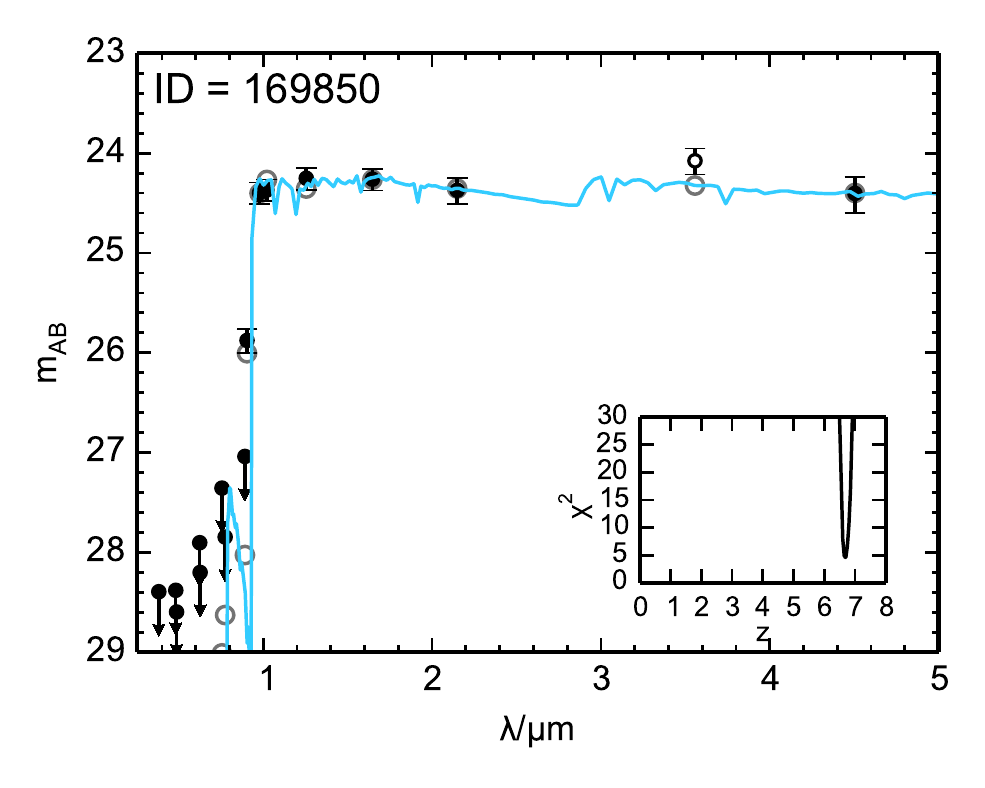}\\
\includegraphics[width = 0.4\textwidth]{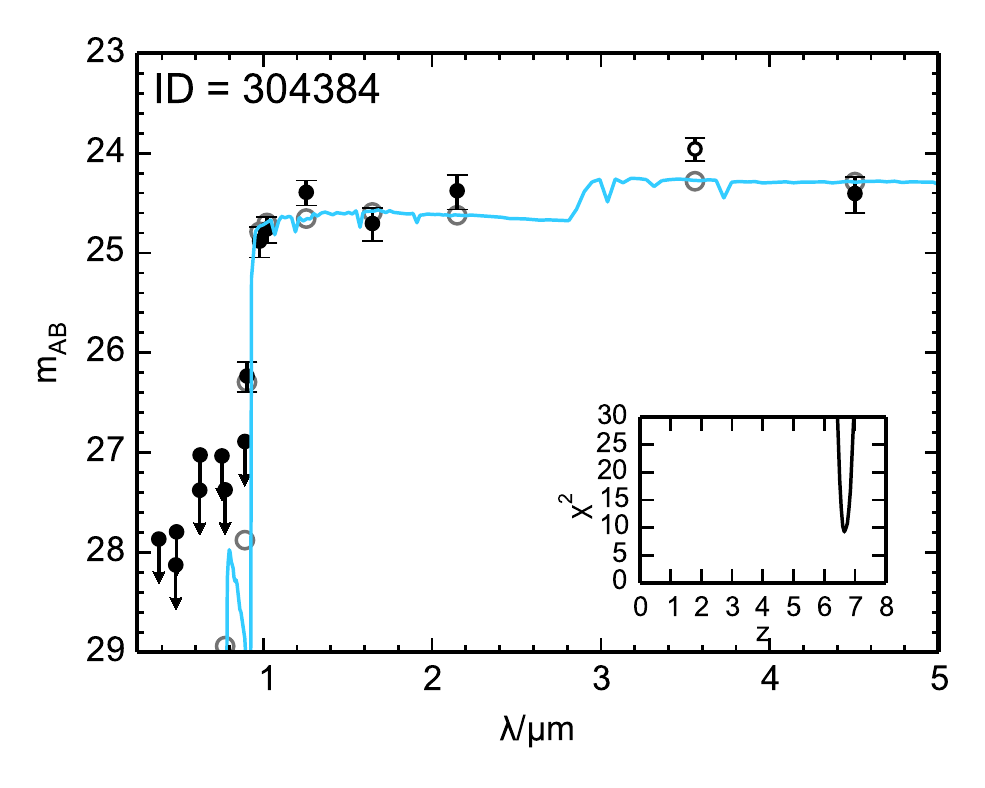}
\includegraphics[width = 0.4\textwidth]{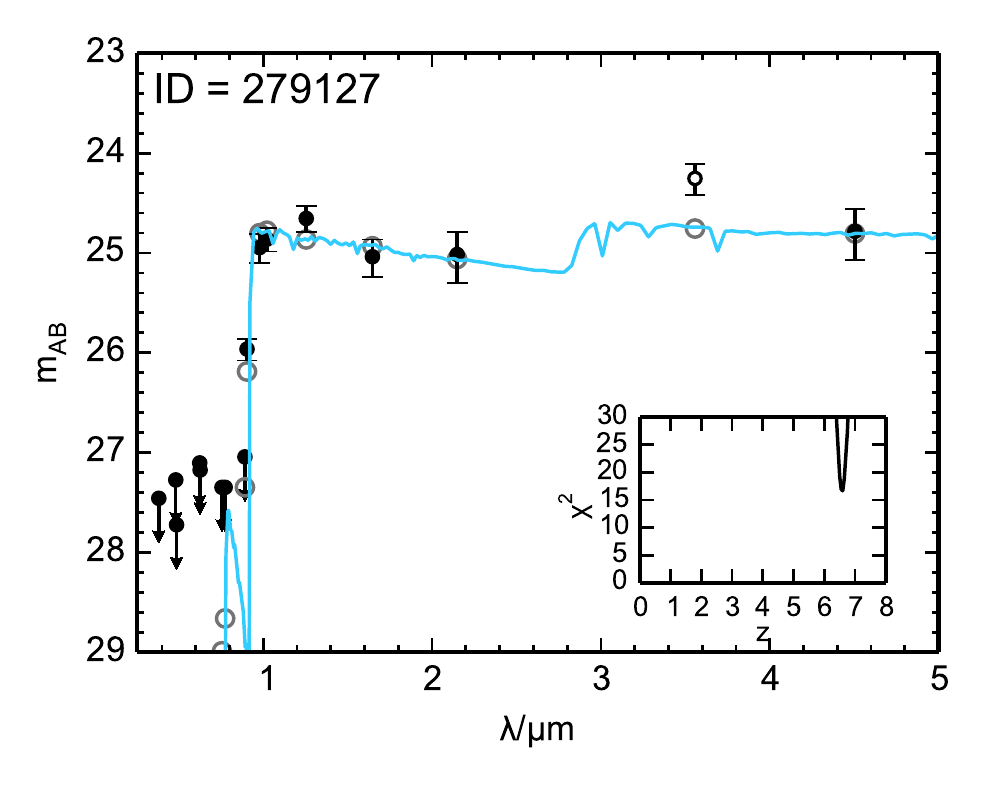}\\

\caption{The optical, near and mid-infrared photometry for the six galaxies targeted with ALMA.
The best-fitting SED model is shown as the blue line.
We excluded either the \chone or \chtwo bands in the fitting due to contamination by rest-frame optical emission lines.
Excluded bands are shown as the open data points.
The grey open circles show the expected magnitude in that band derived from the best-fitting model.
The inset plot shows the $\chi^2$ as a function of redshift derived from our SED fitting analysis.
}\label{fig:sed}

\end{figure*}

In Fig.~\ref{fig:sed} we show the results of our SED fitting analysis to determine the best-fitting photometric redshifts and to provide an estimate of the stellar masses.
We fit to 16 bands, which consists of the CFHT $u^{*}gri$ bands, the Subaru/HSC $GRIZy$ bands, the deeper Subaru/SuprimeCam $z'$-band data~\citep{Furusawa2016}, the VISTA $YJHK_{s}$ bands from UltraVISTA DR3 and finally the available~\emph{Spitzer}/IRAC \chone and \chtwo data.
The inclusion of the HSC $y$-band data in particular results in more precise photometric redshifts for our sample as compared to our previous analyses~\citep{Bowler2014, Bowler2016}, as it probes close to the Lyman-break at $z \simeq 7$, but does not extend as far to the red as the VISTA $Y$-band filter.
In Fig.~\ref{fig:sed} we also show the resulting $\chi^2$ distribution as a function of redshift.
Despite the inclusion of low-redshift contaminant SED models in the fitting proceedure (e.g. old and dusty templates, with up to $A_{V} = 6$), none of the galaxies show an acceptable fit at $z < 6.5$, demonstrating the robustness of the high-redshift solution.
We excluded the~\emph{HST}/WFC3 $JH_{140}$ photometry from the fitting.
While this data is nominally deeper than the ground-based imaging, because the galaxies considered in this study are highly extended, large apertures are required to capture the total flux.
The result is that including the $JH_{140}$ data provides minimal additional constraints on the SED over the narrower ground-based filters (although both datasets are consistent;~\citealp{Bowler2016}).

\section{Investigating potential biases in the measurement of the rest-frame UV slope}
\begin{figure*}

\includegraphics[width = \textwidth]{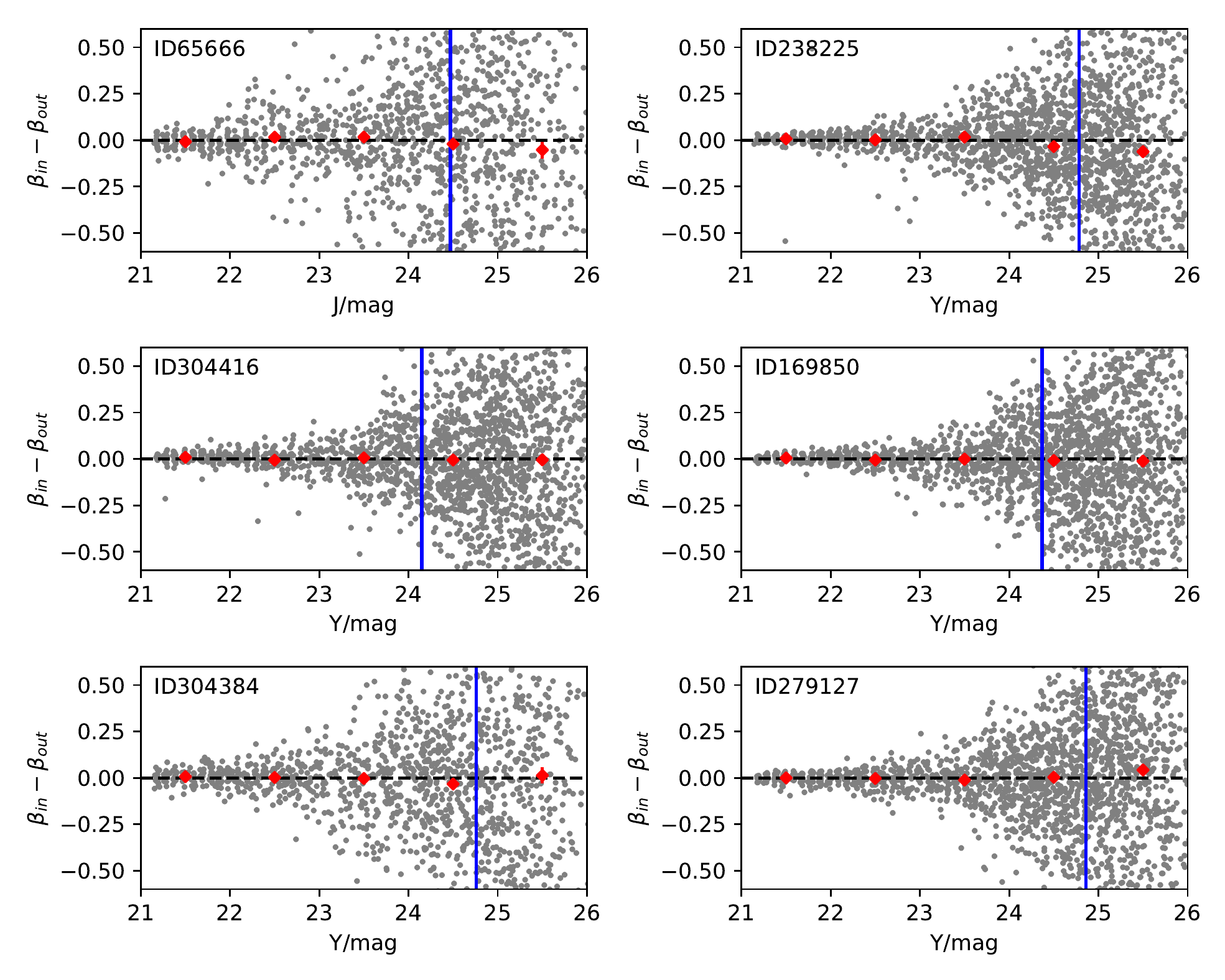}

\caption{The results of injection and recovery simulations with an input of $\beta = -2.0$.
The difference between the input and measured $\beta$ is plotted against the output total magnitude in either the $Y$ or $J$-band.
Results for individual injected sources are shown as the grey points, with the mean and standard error on the mean shown as the red diamonds for bins in magnitude ($\Delta m = 1.0$).
Each panel shows the results of the simulations performed in the region of data surrounding each of our 6 sources.
For ID65666 the results are derived by fitting to the $J$ and $H$ bands only, for the remaining sources the $Y$, $J$ and $H$ are included in the power-law fit to derive $\beta$.
The vertical blue line in each plot shows the observed total magnitude of the object.
A black dashed horizontal line is shown at $\beta_{\rm in} - \beta_{\rm out} = 0.0$ to guide the eye.}\label{fig:betasim}

\end{figure*}

We investigated the robustness of our measurement of $\beta$ using injection and recovery simulations.
Simulated galaxies were created assuming an underlying Sersic profile with a Sersic index of $n = 1.5$ and a half-light radius of $r_{1/2} = 2.0\,{\rm kpc}$.
These parameters were chosen to match those found from stacking the~\emph{HST}/WFC3 imaging for the brightest (and most extended) galaxies at $z \simeq 7$~\citep{Bowler2016}, including the objects in this sample.
We injected these fake sources with apparent magnitudes in the range $[21, 26]$, and with $\beta$ values in the range $[-3.0, 0.0]$.
To account for the varying depth across the UltraVISTA near-IR imaging, simulations were performed in sub-sets of the data ($7\,{\rm arcmin}$ on a side) around each high-redshift candidate.
The simulations exactly reproduced the source extraction and $\beta$ measurements followed for the real objects.
Multiple runs were produced to ensure minimal overlap or overcrowding between simulated sources, which were placed in random blank positions in the image.
Blank positions were determined using the segmentation map produced by {\sc SExtractor}.
In Fig.~\ref{fig:betasim} we show the results of these simulations for an input $\beta = -2.0$.
Approximately $1500$ recovered sources are shown.
We find no significant bias in the recovery of the rest-frame UV slope from the UltraVISTA data, both when fitting to the $Y$, $J$ and $H$ photometry, and also when fitting to the $J$ and $H$ photometry alone (as is the case for ID65666).
We find a larger scatter when only fitting to the $J$ and $H$ bands as expected, however for object ID65666 it is necessary to remove the $Y$ band from the fitting due to the Lyman-break falling in this band.
Finally, the scatter we find agrees well with the derived errors in $\beta$, providing reassurance that these error bars are realistic.

\bsp	
\label{lastpage}
\end{document}